\documentclass[aps,prb,twocolumn,nofootinbib,groupedaddress,floatfix,superscriptaddress,pdflatex]{revtex4-2}

\usepackage{graphics}
\usepackage{graphicx}
\usepackage{amssymb}
\usepackage{amsmath}
\usepackage{amsfonts}
\usepackage{color}
\usepackage{setspace}
\usepackage{multirow}
\usepackage{array}
\usepackage[subrefformat=parens]{subcaption}
\usepackage{url}
\usepackage{hyperref}

\newcommand{\rev}[1]{\textcolor{red}{#1}}
\renewcommand{\rev}[1]{#1} % OFF

\usepackage{siunitx}

\def\PATHFIG{./}

\date{\today}

\begin{document}

\title{Application of Regional Chemical Potential Analysis to Si Adsorption on the Diamond (001) Surface}
%\title{Computational study of Si adsorption on Diamond (001) Surface}

\author{Masahiro Fukuda}
\email{masahiro.fukuda@issp.u-tokyo.ac.jp}
\affiliation{Institute for Solid State Physics, The University of Tokyo, 5-1-5 Kashiwanoha, Kashiwa, Chiba 277-8581, Japan}

\author{Arath E. Marin Ramirez}
\affiliation{Institute for Solid State Physics, The University of Tokyo, 5-1-5 Kashiwanoha, Kashiwa, Chiba 277-8581, Japan}

\author{Yoshiaki Sugimoto}
\affiliation{Department of Advanced Materials Science, The University of Tokyo, 5-1-5 Kashiwanoha, Kashiwa, Chiba 277-8561, Japan}

\author{Taisuke Ozaki}
\affiliation{Institute for Solid State Physics, The University of Tokyo, 5-1-5 Kashiwanoha, Kashiwa, Chiba 277-8581, Japan}

\begin{abstract}
    Adsorption of carbon dimers and Si atoms on the reconstructed diamond (001) surface is investigated using density functional theory and regional chemical potential (RCP) analysis. We first demonstrate that the RCP distribution provides a real-space description of the bonding rearrangements responsible for the site-selective growth of experimentally observed carbon-dimer ribbons. We then examine the adsorption of single and multiple Si atoms. The calculated adsorption energies show that a single Si atom preferentially bridges a surface carbon dimer and that subsequently adsorbed Si atoms favor neighboring dimer sites through Si-Si bond formation. The RCP analysis identifies electron-donating regions at the ends of finite Si chains, providing an intuitive explanation for their preferential one-dimensional growth and a physically motivated strategy for selecting candidate adsorption structures. At higher Si coverages, geometry optimizations yield Si stripe and planar square-lattice structures on the diamond surface. Surface phase analysis indicates that an increase in the effective Si chemical potential favors structures with progressively higher Si coverages, from the Si stripe phase to the Si square-lattice phase. The calculated band structures reveal a progressive reduction of the surface band gap with increasing Si coverage.
    In the Si square-lattice structure, several bands cross the Fermi level, and the conducting states along the two in-plane directions have distinct Si and C character because the underlying diamond (001) substrate lacks fourfold rotational symmetry.
    These results establish RCP analysis as a useful approach for interpreting surface covalent bonding and guiding the exploration of adsorption-driven surface structures.

\end{abstract}

    \maketitle

    %%%%%%%%%%%%%%%%%%%%%%%%%%%%%%%%%%
    \section{Introduction}
    %%%%%%%%%%%%%%%%%%%%%%%%%%%%%%%%%%

    %Surface doping of diamond surfaces is a promising method to control the electronic properties of diamond films
    %for applications in electronics and optoelectronics. \cite{yangProgressStructuralElectronic2022} The effect of Silicon doping
    %has been studied on carbon defects on nano-diamond films to facilitate crystal growth.
    %Recently, high resolution near-contact AFM techniques allow the observation of ribbon structures on clean diamond surfaces. This allows us to experimentally
    %observe the growth mechanism of Si doping in diamond films.
    %This has been done with C atoms, and Si atoms provide a possibility of enhancing properties of the clean diamond film. The RCP method has been
    %developed in order to visualize possible bonding sites, to accelerate the finding of sites of preferential electron attachment.
    %Previous studies have been conducted on (100) diamond surface
    %\cite{schenkFormationSiliconTerminated2015}. Previous computational studies
    %have been conducted showing the behavior of single Si atoms deposited within
    %(001) surface within carbon islands \cite{renEvolutionBehaviorSi2015}.
    %However, as to our knowledge, thus far no studies have been conducted for Si
    %adsorption on the (001) for a clean diamond surface. Recently, AFM studies
    %have been conducted as to image the  formation of ribbon dimers in clean diamond
    %(001) surfaces \cite{zhangDimerRibbonStructures2025}. Motivated by this,
    %we propose a computational study of Si adsorption on clean (001) diamond surface
    %using density functional theory (DFT).

\begin{figure}[h]
        \centering
        \includegraphics[width=0.45\textwidth]{\PATHFIG/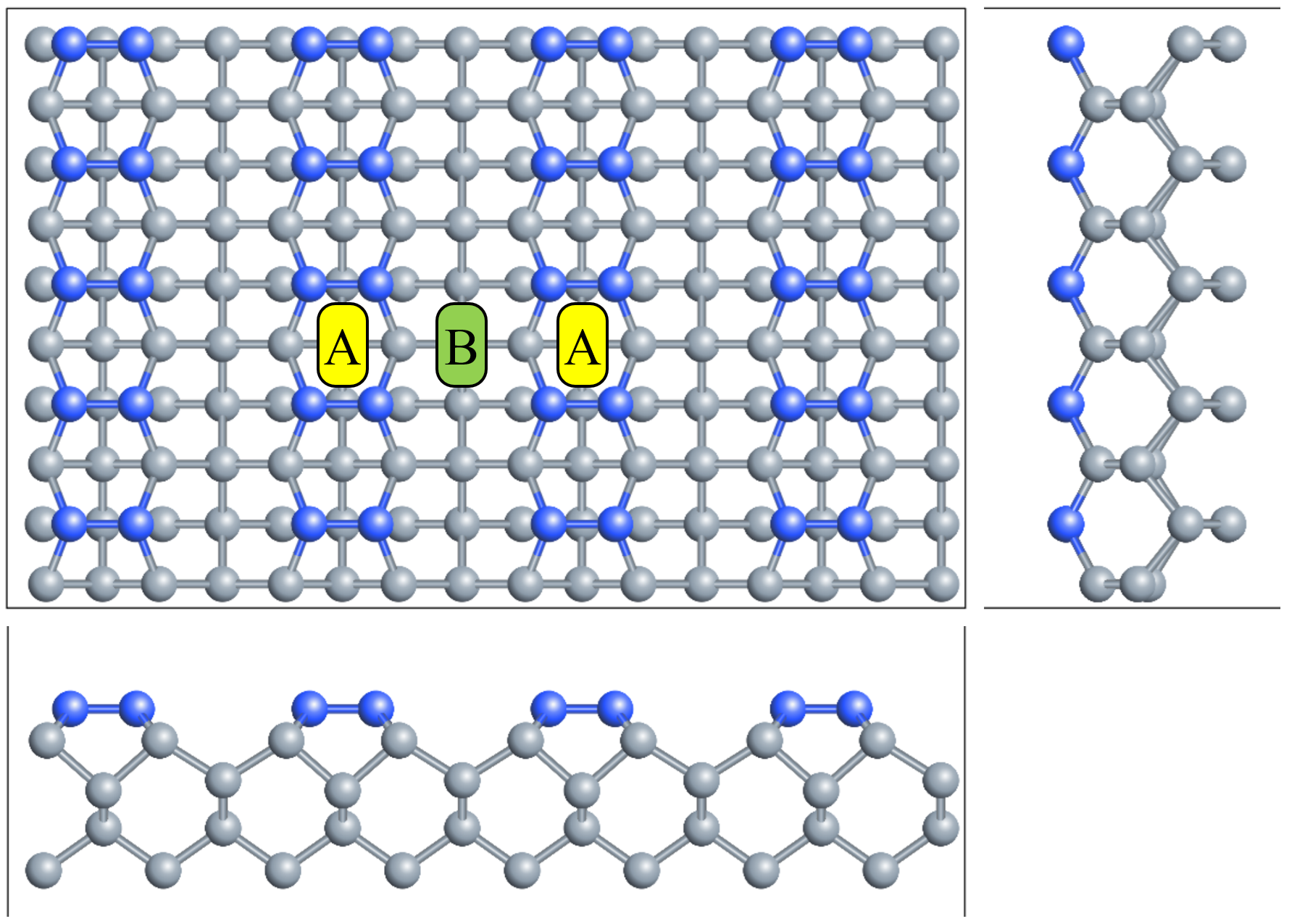}
        \caption{Structure of the $4 \times 5$ dimerized diamond (001) surface. The A and B sites
        denote possible adsorption sites for carbon dimers.
        \rev{The carbon atoms constituting the surface dimer are depicted in blue.}
        }
        \label{fig:structures_C_ribbons}
    \end{figure}

    Diamond films have attracted significant attention as electronic device
    materials~\cite{10.1098/rsta.2023.0382}, and substantial progress has been made in identifying surface structures
    through experiments as well as analyzing electronic properties using first-principles
    calculations~\cite{Ristein2006,PhysRevB.51.14669}.
    Recently, high-resolution near-contact atomic force microscopy
    (AFM) has enabled the observation of ribbon structures on non-hydrogen-terminated diamond surfaces at the atomic scale.
    Previous AFM studies have demonstrated the formation of carbon dimer ribbons on the diamond (001) surface~\cite{Zhang2025,PhysRevResearch.7.023036}.
    Furthermore, discussions based on the adsorption energy of dimers have
    revealed that dimers preferentially adsorb at specific sites on the diamond (001)
    surface. In particular, Density Functional Theory (DFT) calculations have
    revealed that the A, AA, and ABA configurations of the dimer ribbons (as illustrated
    in Fig.~\ref{fig:structures_C_ribbons}) are energetically stable, and
    these structures have been experimentally observed by AFM~\cite{PhysRevResearch.7.023036}.
 
    In this study, to clarify the origin of this site selectivity,
    we employ the Regional Chemical Potential (RCP) analysis method proposed in Ref.~\cite{10.1063/5.0288934},
    which evaluates the covalent bond-forming ability of material surfaces.
    Using this RCP analysis, we visualize chemically
    reactive sites where carbon ad-dimers preferentially adsorb, thereby elucidating
    the growth mechanism of the surface structure.

    Although the above study focuses on the carbon dimer adsorption process on diamond
    surfaces, Si doping is also important for controlling the
    electronic properties of diamond in practical applications~\cite{CHEN20123021,yangProgressStructuralElectronic2022}.
    The structural and electronic properties of SiC and Si/diamond interfaces have also attracted considerable attention because of their potential applications in electronic devices~\cite{EDHOLM1997245,10.1021/acsaelm.0c00289,10.1063/5.0270223}.
    Recently, STM experiments combined with DFT calculations have revealed the adsorption behavior of carbon dimers on the Si(001) surface~\cite{cowie2026atomicallyprecisemechanosynthesiscarbon}.
    In contrast, the adsorption of Si atoms on the diamond (001) surface remains largely unexplored, despite its fundamental interest.
    Although the adsorption of a single Si atom on the diamond (001) surface has been investigated using DFT~\cite{renEvolutionBehaviorSi2015},
    the adsorption of multiple Si atoms and the associated adsorption mechanisms have not yet been theoretically elucidated.

    %The effects of silicon doping on carbon defects in nanodiamond films have been
    %studied to facilitate crystal growth, enabling experimental observation of
    %the growth mechanisms associated with Si incorporation~\cite{}.

    In this study, we further aim to elucidate the adsorption mechanism of multiple Si atoms on a diamond (001) surface by DFT calculations and RCP analysis.
    As discussed later, this analysis predicts that Si stripe structures and a Si square lattice can be formed on the diamond (001) surface through the adsorption of multiple Si atoms.
    %Their electronic conduction characteristics are then investigated and discussed based on detailed band structure analysis.
    \rev{
    Their electronic conduction characteristics and stability are also investigated through band-structure analyses and a surface phase diagram.
    }
    
    \rev{
    This paper is organized as follows. Section \ref{sec:computational_details} describes the computational methods and models employed in the DFT calculations and RCP analysis.
    Section \ref{sec:C001} applies RCP analysis to carbon-dimer adsorption on the diamond (001) surface and discusses the formation mechanism of carbon-dimer ribbons.
    Sections \ref{sec:Si_C001_energy} and \ref{sec:Si_C001_RCP} examine the adsorption energetics of single and multiple Si atoms and elucidate their adsorption selectivity and chain-growth mechanisms using RCP analysis.
    Section \ref{sec:Si_stripe_square_lattice} presents the Si stripe and square-lattice structures obtained at higher Si coverages, and Sec.~\ref{sec:band_analysis} discusses their electronic structures and conduction characteristics.
    Section \ref{sec:phase_transition} analyzes the relative stability of the Si-covered surface phases as a function of the Si chemical potential.
    Finally, Sec.~\ref{sec:conclusion} summarizes the main conclusions of this study.
    }

    %%%%%%%%%%%%%%%%%%%%%%%%%%%%%%%%%%
    \section{Computational Details} \label{sec:computational_details}
    %%%%%%%%%%%%%%%%%%%%%%%%%%%%%%%%%%
    %Computations were carried out using the OpenMX package~\cite{ozakiOpenmxSoftwarePackage2004,ozakiEfficientFirstprinciples2003}.
    %The exchange-correlation functional was treated within the GGA-PBE scheme.
    %Regional chemical potential analysis  \cite{fukudaRegionalChemicalPotential2025} was performed as the main means
    %to analyze the most active sites within an Si structure.

    The DFT calculations within a generalized gradient approximation (GGA)
    \cite{PhysRev.140.A1133,PhysRevLett.77.3865} were performed for the geometry
    optimizations using the OpenMX code~\cite{OpenMX} based on norm-conserving
    pseudopotentials generated with multireference energies
    \cite{PhysRevB.47.6728} and optimized pseudoatomic basis functions~\cite{PhysRevB.67.155108}.
    For each C atom, two, two, and one optimized radial functions were allocated
    for the $s$, $p$, and $d$ orbitals,
    respectively, 
    corresponding to the s2p2d1 basis specification.
    The s2p2d1 basis specification was also adopted for Si atoms.
    A cutoff radius of 6 bohr
    was chosen for the basis functions of C atoms, whereas 7 bohr was chosen for
    Si atoms. The qualities of basis functions and fully relativistic
    pseudopotentials were carefully benchmarked by the delta gauge method~\cite{Lejaeghere2016-gh}
    to ensure accuracy of our calculations. An electronic temperature of 300 K was
    used to count the number of electrons by the Fermi-Dirac function. 
    A real-space grid corresponding to an energy cutoff of 220 Ry was used for the numerical integration
    and for the solution of the Poisson equation~\cite{PhysRevB.72.045121}.
    The geometry optimizations were performed using a combination scheme of the rational function
    (RF) method~\cite{doi:10.1021/j100247a015} and the 
    direct inversion in the iterative subspace (DIIS) method~\cite{CSASZAR198431} with a BFGS update~\cite{10.1093/imamat/6.1.76,
    10.1093/comjnl/13.3.317, 10.2307/2004873, 10.2307/2004840} for the approximate
    Hessian. The threshold of the forces for geometry optimizations was set to
    be 0.0003 Hartree/bohr.
    All the calculations were performed without spin polarization.
    Only the $\Gamma$ point was used for geometry optimizations, except for the Si stripe structure and Si square lattice in Secs.~\ref{sec:Si_stripe_square_lattice} and \ref{sec:band_analysis}.

    The slab models used in this study were constructed using supercells generated from
    the optimized diamond (001)-(2×1) surface with in-plane lattice-vector lengths of 5.052 \AA\ $\times$ 2.526 \AA\
    in the same manner as the previous studies of Refs.~\cite{Zhang2025,PhysRevResearch.7.023036}.
    The bottom two atomic layers of carbon atoms of the substrate were fixed during the
    geometry optimization.
    A vacuum region thicker than 20 \AA\ was introduced to suppress interactions between periodic images of the slab.

    RCP analysis~\cite{10.1063/5.0288934}
    was performed to analyze the most
    reactive sites on material surfaces. Calculations of the local RCPs and the kinetic
    energy density were performed by FLPQ module~\cite{FLPQ} in QEDalpha package~\cite{QEDalpha}.
    %Local quantities were evaluated using the energy range between -2.0 eV and 0.0 eV.
    %OpenMX Viewer~\cite{LEE2019192} and FLPQViewer~\cite{FLPQViewer} were used
    %for visualization of the atomic structures and RCPs.
    Atomic units were used
    for the RCP.\footnote{ The unit of the RCP in Ref.~\cite{10.1063/5.0288934} was erroneously given as Hartree/Bohr$^3$. The correct unit is Hartree.}
    The broadening
    factors for RCP calculations were set to $\tilde{T}_{\rm lower}= \tilde{T}_{\rm
    upper}= 0.001 / k_{B}[{\rm K}]$ for all the systems, where the Boltzmann
    constant $k_{B}=8.617 \times 10^{-5}[{\rm eV}/{\rm K}]$.
    The energy range from $-2$\,eV to 0\,eV was chosen for the RCP calculations to approximately correspond to the highest occupied molecular orbital (HOMO) energy levels of C and Si atoms, which are responsible for covalent bond formation.

    For the band calculations for the Si stripe structure and the Si square lattice discussed in Sec.~\ref{sec:band_analysis} and surface phase analysis in  Sec.~\ref{sec:phase_transition},  a $15 \times 11 \times 1$ mesh of k points was used to improve the accuracy. The same k-point mesh was also used for the geometry optimization.

    For the Closest Wannier Function (CWF) calculations~\cite{PhysRevB.110.125115} for the Si square lattice,
    the Fermi window function was adopted with a $31 \times 21 \times 1$ mesh of k points. 
    The energy window ranging from $-25.0$ eV to 0.0 eV was used, with smearing parameters of 0.1 and 1.0 eV at the lower and upper edges of the window, respectively.

    The structure and CWFs were visualized using OpenMX Viewer~\cite{LEE2019192}, the RCPs were visualized using FLPQViewer~\cite{FLPQViewer}, and the band structures were visualized using PQViewer~\cite{PQViewer}.
    
    %%%%%%%%%%%%%%%%%%%%%%%%%%%%%%%%%%
    \section{Results and Discussion} \label{sec:Result}
    %%%%%%%%%%%%%%%%%%%%%%%%%%%%%%%%%%

    %%%%%%%%%%%%%%%%%%%%%%%%%%%%%%%%%%
    \subsection{Formation of C dimer ribbons on the diamond (001) surface} \label{sec:C001}
    %%%%%%%%%%%%%%%%%%%%%%%%%%%%%%%%%%

    %-- more explanation about ribbon structure --

    On the diamond (001) surface with a (2×1) reconstruction, carbon adatoms are
    stabilized via the formation of paired structures, known as ad-dimers. In
    Ref.~\cite{PhysRevResearch.7.023036}, as one, two, and three carbon ad-dimers are
    sequentially added onto the clean diamond (001) surface shown in Fig.~\ref{fig:structures_C_ribbons},
    the A, AA, and ABA configurations are obtained as the most stable structures through geometry
    optimization.
    In this section, the adsorption mechanism of the carbon ad-dimer is elucidated
    through RCP analysis~\cite{10.1063/5.0288934}.

    RCP analysis visualizes electron-donating regions on material surfaces that tend to form covalent bonds, thereby enabling the identification of preferential adsorption sites involving covalent bond formation.
    Specifically, the RCP is visualized by plotting its values integrated over the energy range from the Fermi level (0~eV) to $-$2~eV, which corresponds to the energy range relevant to covalent-bond formation, on the material surface defined by the zero isosurface of the kinetic energy density.
    The kinetic energy density of the electrons in the non-relativistic framework is defined as~\cite{Tachibana2001}
    {\small
    \begin{align}
        {T}_e (\vec{r}) \equiv -\frac{\hbar^2}{4m_e} \left[ {\psi}^\dagger (\vec{r}) \nabla^2 {\psi}(\vec{r}) \right] + h.c., 
    \end{align}
    }%
    where $m_e$ is the electron mass.
    The local  RCP~\cite{Tachibana1992,tachibana1999,Tachibana2001,10.1063/1.3072369,10.1063/1.3651182,10.1063/1.2973634}, $\mu_R^{\tau}(\vec{r}) \equiv
    \Delta \varepsilon_{\tau}(\vec{r}) / \Delta n(\vec{r})$,
    is defined as the change in the regional energy density
    $\varepsilon_{\tau}(\vec{r})$ with respect to the change in the
    electron number density $n(\vec{r})$.
    At the material surface, the regional energy density is given by
    $\varepsilon_{\tau}(\mathrm{surface}) = -\frac{\hbar^2}{8m_e} \nabla^2 n(\mathrm{surface}).$
    Since $\nabla^2 n (\vec{r})$ serves as a measure of electron delocalization,
    the RCP at the material surface can be interpreted as the response of
    electron delocalization to changes in the electron number density.
    Regions with large RCP values exhibit a high electron-donating ability and are therefore more likely to contribute to covalent-bond formation.
    Particularly large RCP values are observed around double bonds and dangling bonds.

\begin{figure*}[htpb]
        \centering
        \includegraphics[width=0.90\textwidth]{\PATHFIG/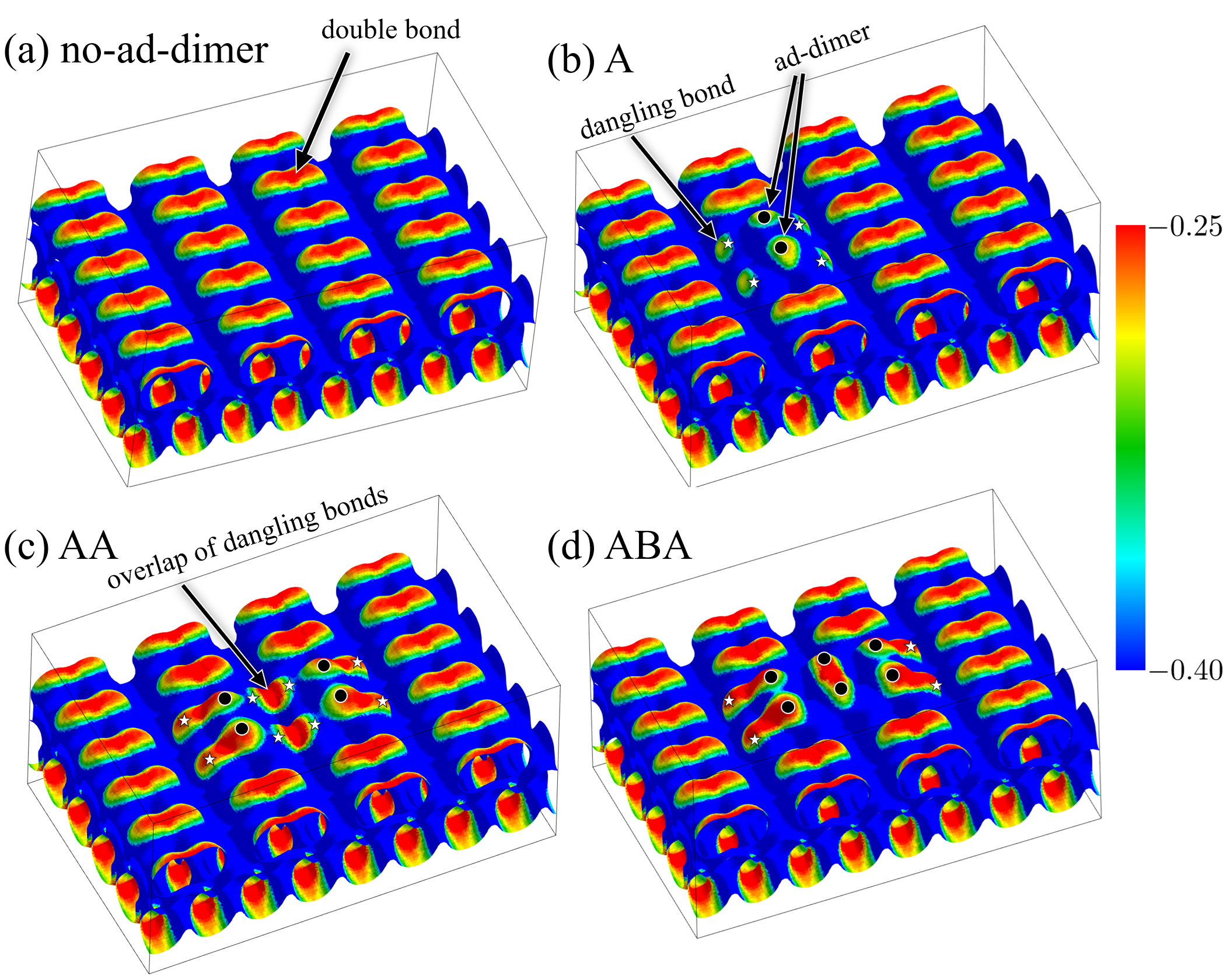}

        \caption{The RCPs on the surface for (a) the no-ad-dimer, (b) A, (c) AA, and
        (d) ABA configurations of carbon ad-dimers on the diamond (001) surface.
        Black circles represent the positions of the ad-dimer atoms, and white stars indicate the positions of the C atoms that constitute the adsorption sites occupied by the ad-dimers.}
        \label{fig:RCP_C_ribbons}
    \end{figure*}

    First, we demonstrate an application of the RCP analysis using the clean diamond (001) surface as
    an example. In Fig.~\ref{fig:RCP_C_ribbons}, the RCP is plotted on the zero
    isosurface of the kinetic energy density, which represents the material surface.
    \rev{As shown in Fig.~\ref{fig:RCP_C_ribbons} (a)} the magnitude of the RCP reflects the degree of electron-donating character,
    and the red regions indicate C–C double-bond–like regions with high electron-donating
    ability. Indeed, a comparison of the C–C bond lengths in the surface dimers
    and in bulk diamond shows values of 1.40 \AA\ and 1.55 \AA, respectively,
    indicating that the carbon dimers have bond lengths closer to those of double
    bonds.

    On the basis of the above findings, we further analyze the bonding configurations of
    the A, AA, and ABA structures using the RCP analysis. 
    \rev{
    In A structure shown in Fig.~\ref{fig:RCP_C_ribbons}(b), high-RCP regions, which can be regarded as dangling bonds, are observed on the four carbon atoms surrounding the adsorption site of the carbon ad-dimer. These four carbon atoms are indicated by white stars in Fig.~\ref{fig:RCP_C_ribbons}(b).
    }
    No significant high-RCP regions
    are found between these carbon atoms, indicating that their bonding state differs
    from that of a double bond.
    %The C–C bond lengths are in the range of 1.44 \AA\ to 1.49
    %\AA, which are intermediate between those of typical single and double bonds.
    \rev{
    The C–C bond lengths range from 1.44 to 1.49 \AA, lying between those of typical single and double bonds (1.55 \AA\ in bulk diamond and 1.34 \AA\ in ethylene, respectively).
    }

    \rev{For the structure with two ad-dimers as shown in Fig.~\ref{fig:RCP_C_ribbons} (c)}, the AA configuration is known
    from previous studies to be energetically favorable. This is because adsorption
    at the A site provides a larger energy gain, and when ad-dimers occupy adjacent
    A sites, the dangling bonds present in the A structure tend to form bonds with
    each other, leading to further stabilization.
    \rev{
    The RCP analysis in Fig.~\ref{fig:RCP_C_ribbons} (c)
    indicates that four carbon atoms located at the center of the ribbon,
    which form the dimer adsorption site B, tend to form two C-C bonds.
    }
    In contrast, the adatoms at both
    ends of the ribbon form double bonds with the underlying substrate carbon atoms, \rev{which are indicated by the white stars}.

    When considering the adsorption of a third carbon dimer, the most stable
    configuration is achieved when the additional dimer bonds with the four
    dangling bonds present in the AA structure. In this ABA structure, the
    central ad-dimer forms a double bond, while the atoms in the two terminal ad-dimers
    form double bonds with the substrate carbon atoms indicated by white stars in Fig.~\ref{fig:RCP_C_ribbons} (d), similar to the AA
    structure.

    As demonstrated above, the purpose of the RCP analysis is not merely to interpret structures obtained after geometry optimization. Rather, the RCP distribution is used to identify electron-donating surface regions that are likely to participate in covalent-bond formation, thereby reducing the configurational search space for subsequently adsorbed atoms. 
    RCP analysis can reveal the chemical bonding characteristics of complex surfaces that cannot be inferred from atomic configurations alone, thereby facilitating the identification of candidate structures and supporting the results obtained from geometry optimization and total-energy comparisons.

    %%%%%%%%%%%%%%%%%%%%%%%%%%%%%%%%%%%%%%%%%%%%%%
    %\subsection{Energetic comparison of Si adsorption sites}
    \subsection{Adsorption energies of Si atoms on the diamond (001) surface} \label{sec:Si_C001_energy}
    %%%%%%%%%%%%%%%%%%%%%%%%%%%%%%%%%%%%%%%%%%%%%%

\begin{figure}[h]
        \centering
        \includegraphics[width=0.3\textwidth]{
            \PATHFIG/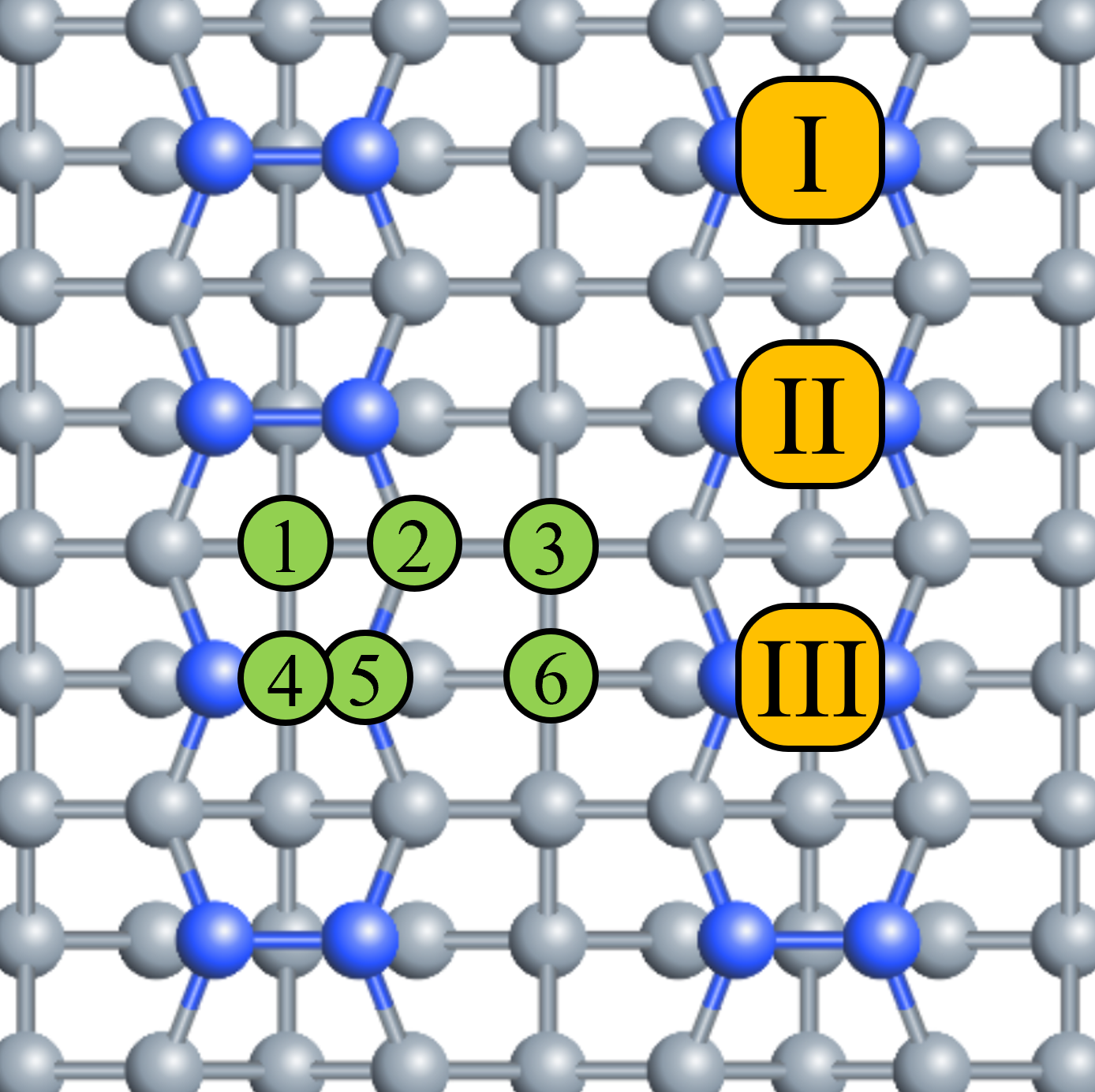
        }
        \caption{Candidate adsorption sites for Si on the diamond (001) surface.}
        \label{fig:adsorption_sites_C001}
    \end{figure}

    To understand the basic behavior of adsorption of a single Si atom on a clean diamond (001) surface, geometry optimizations were performed for a single Si adatom on the $4 \times 5$ diamond (001) surface shown in Fig.~\ref{fig:structures_C_ribbons}.
    For the case of C ad-dimers discussed in the previous subsection, experiments have
    shown that C adatoms form ribbons on the diamond (001) surface as dimers rather
    than as single atoms. In contrast, for Si adsorption, considering the
    larger atomic radius of Si relative to the reactive sites, we assume that Si
    atoms adsorb individually and analyze the adsorption configurations based on this assumption.
    We placed a single Si atom at six sites shown in Fig.~\ref{fig:adsorption_sites_C001} as initial configurations
    and performed geometry optimization. The adsorption energies, defined as $E_{\mathrm{ads}}
    = E_{\mathrm{C(001)}+n\mathrm{Si}}- E_{\mathrm{C(001)}}- n E_{\mathrm{Si\ atom}}$,
    obtained from the optimized structures are summarized in Table ~\ref{tab:FE_Si}.
    Initial configurations 4 and 5 relaxed to the same lowest-energy adsorption structure, in which the Si atom bridges a surface C dimer.

    %Formation energies, defined as $E_{\mathrm{form}} = E_{\mathrm{Si}} - E_{\mathrm{noSi}} - n E_{\mathrm{abs}}(\mathrm{Si})$ were
    %calculated for each configuration and the most stable configuration was selected as the
    %basic unit of Si adsorption.
    %A periodic slab of To find the basic unit of adsorption, symmetrical sites were selected for the single Si atom based on \cite{schenkFormationSiliconTerminated2015}.

    \begin{table}[h]
        \caption{Adsorption energies of a single Si atom on the diamond (001) surface.}
        \label{tab:FE_Si}
        \centering
        \begin{tabular}{c c S}
            \hline
            {Initial site} & {Final site} &  {$E_{\mathrm{ads}}$ (eV)} \\
            \hline
            1    &  1    & -4.463    \\
            2    &  2    & -4.428    \\
            3    &  2    & -4.428    \\
            4    &  4    & -5.126    \\
            5    &  4    & -5.126    \\
            6    &  6    & -4.191    \\
            \hline
        \end{tabular}
    \end{table}

    \begin{table}[h]
        \centering
        \caption{Adsorption energies of multiple Si atoms on the diamond (001) surface.
        The energy $E_{\mathrm{ads}}$(I) represents $E_{\mathrm{ads}}$ when the Si atom is located at site I. }
        \label{tab:FE_multi_Si}
        \begin{tabular}{c S S c}
            \hline
            {Si site}   & {$E_{\mathrm{ads}}$ (eV)} & {$E_{\mathrm{ads}}$ - $nE_{\mathrm{ads}}$(I) (eV) } \\
            \hline
            %0          & 0.000    & {---}    \\
            I           &  -5.128                 & {---}                  \\
            I,II        & -11.045                 & -0.790                 \\
            I,III       & -10.220                 & 0.035                  \\
            I,II,III    & -17.527                 & -2.143                 \\
            %I,II,III,IV & -416.127                   & -3.412                 \\
            \hline
        \end{tabular}
    \end{table}

    %\begin{table}[h]
    %\centering
    %\begin{tabular}{c c c c}
    %\hline
    %{Si site} & {$E_{\mathrm{form}}$ (eV)} & {$E_{\mathrm{form}}$ - $E_{\mathrm{form}}$(0) (eV) } \\
    %\hline
    %%0      & 0.000   & {---}    \\
    %1       & -103.179 & {---}   \\
    %1,2     & -207.147 & -0.790  \\
    %1,3     & -206.322 & 0.035   \\
    %1,2,3   & -311.680 & -2.143  \\
    %1,2,3,4 & -416.127 & -3.412  \\
    %\hline
    %\end{tabular}
    %\label{tab:FE_multi_Si}
    %\end{table}

    %\begin{table}[h]
    %\centering
    %\begin{tabular}{c c c c}
    %\hline
    %Position & Formation energy (eV) & \\
    %\hline
    %0 & -31555.235 & 0 &  \\
    %1 & -31666.499 & -111.264 &  \\
    %1,2 & -31778.553 & -223.318 & -0.78951 \\
    %1,3 & -31777.728 & -222.493 & 0.03545 \\
    %1,2,3 & -31891.171 & -335.936 & -2.14323 \\
    %1,2,3,4 & -32003.704 & -448.469 & -3.41199 \\
    %\hline
    %\end{tabular}
    %\label{tab:FE_multi_Si}
    %\end{table}

    %\rev{
    %From the basic unit of Si adsorption, more Si atoms were systematically
    %placed in the different sites of the surface for 2, 4 and 6 Si atoms in order
    %to find a preferred adsorption configuration. Similarly to the single atom
    %case, the adsorption energies are calculated until the most stable one is found.
    %}

    Based on the discussion of the adsorption energy of a single Si atom, which indicates that Si atoms preferentially adsorb on carbon dimers, we next consider structures in which multiple Si atoms are adsorbed on carbon dimers.
    The adsorption energies for one to three Si atoms adsorbed at sites I–III in Fig.~\ref{fig:adsorption_sites_C001} are summarized in Table ~\ref{tab:FE_multi_Si}.
    Table ~\ref{tab:FE_multi_Si} shows that adsorption at adjacent sites I and II is energetically more favorable than adsorption at separated sites I and III.
    The results indicate that, when multiple Si atoms are adsorbed, they tend to occupy neighboring dimers, suggesting that the formation of a Si chain along the dimer-chain direction is energetically favorable.
    As a result, it was found that a linear Si wire structure, in which Si atoms are continuously adsorbed above each C dimer, is energetically favored.
    This behavior contrasts with the case of C dimer adsorption, in which a C dimer ribbon formed by adsorption at every other site in the direction perpendicular to the dimer chain was energetically preferred.

    %%%%%%%%%%%%%%%%%%%%%%%%%%%%%%%%%%%
    \subsection{RCP analysis of Si adsorption on the diamond (001) surface} \label{sec:Si_C001_RCP}
    %%%%%%%%%%%%%%%%%%%%%%%%%%%%%%%%%%%

\begin{figure}[h]
        \centering
        \includegraphics[width=0.45\textwidth]{\PATHFIG/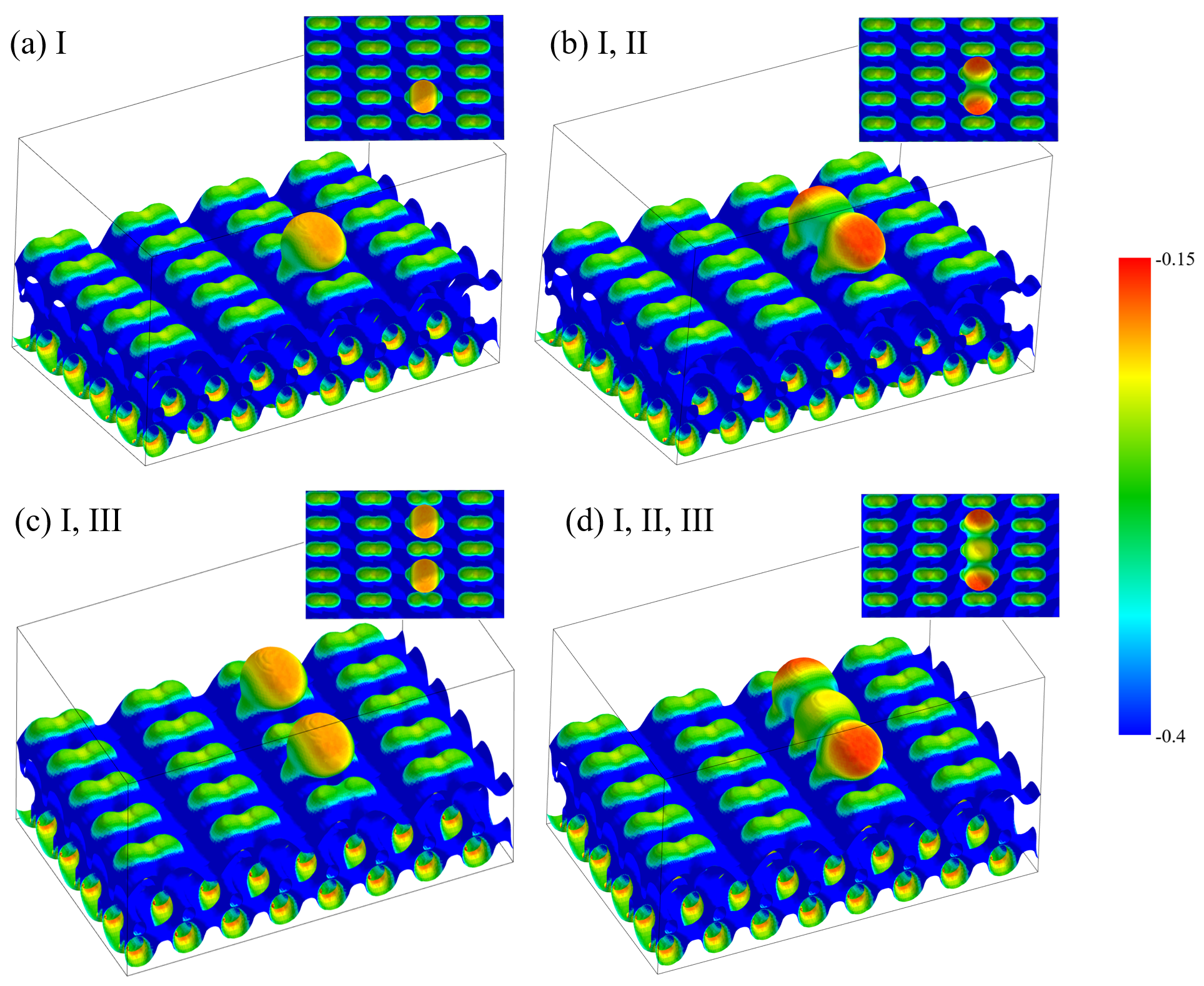}
        \caption{RCP analysis for the Si adsorption on the diamond (001) surface.}
        \label{fig:RCP_Si_addatoms}
    \end{figure}

    The RCP analysis provides a more intuitive understanding of the adsorption process discussed in terms of adsorption energy. Figure \ref{fig:RCP_Si_addatoms} shows the RCP analysis results for one to three Si atoms adsorbed on the diamond (001) surface.
    First, Fig.~\ref{fig:RCP_Si_addatoms} (a) indicates that the atomic size of Si is larger than the interatomic distance of the C dimer.
    Therefore, a Si atom is expected to be more stable when adsorbed on both C atoms of a dimer rather than bonded to only one C atom.
    Comparing Figs.~\ref{fig:RCP_Si_addatoms} (b) and (c), which correspond to the adsorption of two Si atoms, it can be seen that adsorption on adjacent dimer sites is energetically favorable because the Si atoms form a bond with each other, leading to stabilization.
    Furthermore, the two bonded Si atoms exhibit red regions at both ends, indicating high RCP values, i.e., dangling bonds with strong electron-donating character.
    This suggests that additional Si atoms can preferentially bond to these dangling bonds when further adsorption occurs.
    This suggestion is supported by the adsorption result of the third Si atom shown in Fig.~\ref{fig:RCP_Si_addatoms} (d).
    Consequently, newly adsorbed Si atoms are expected to attach to the terminal Si atoms, providing an intuitive explanation for the preferential growth of a Si chain along the dimer-chain direction.

    While adsorption energy analysis determines stable structures through geometry optimization starting from many initial configurations, RCP analysis offers the advantage of predicting and narrowing down candidate initial configurations that are likely to evolve into stable adsorption structures.
    
    %RCP analysis was performed on the single atom configuration to observe the most
    %active sites.
%
    %From this behavior several configurations were constructed to find the most energetically
    %favorable position within DFT.
    %Out of all configurations, the 2x1
    %reconstruction with Si atoms on top of the dimer rows was found to be the
    %most energetically favorable configuration.

    %%%%%%%%%%%%%%%%%%%%%%%%%%%%%%%%%%%
    \subsection{Si stripe structure and Si square lattice} \label{sec:Si_stripe_square_lattice}
    %%%%%%%%%%%%%%%%%%%%%%%%%%%%%%%%%%%

    When Si atoms are adsorbed on top of all dimers of the diamond (001) surface, geometry optimization yields a stripe structure consisting of aligned Si chains. Furthermore, when the gaps between the stripes are completely filled with additional Si atoms, geometry optimization results in a stable configuration in which the Si atoms form a square lattice. Since Si surface structures may exhibit complex buckling patterns, other configurations could potentially be obtained by selecting different initial structures. In the present study, however, we focus on the physical properties of the obtained Si stripe structure and Si square lattice.
    
\begin{figure}[htbp]
        \centering
        \includegraphics[width=0.45\textwidth]{
            \PATHFIG/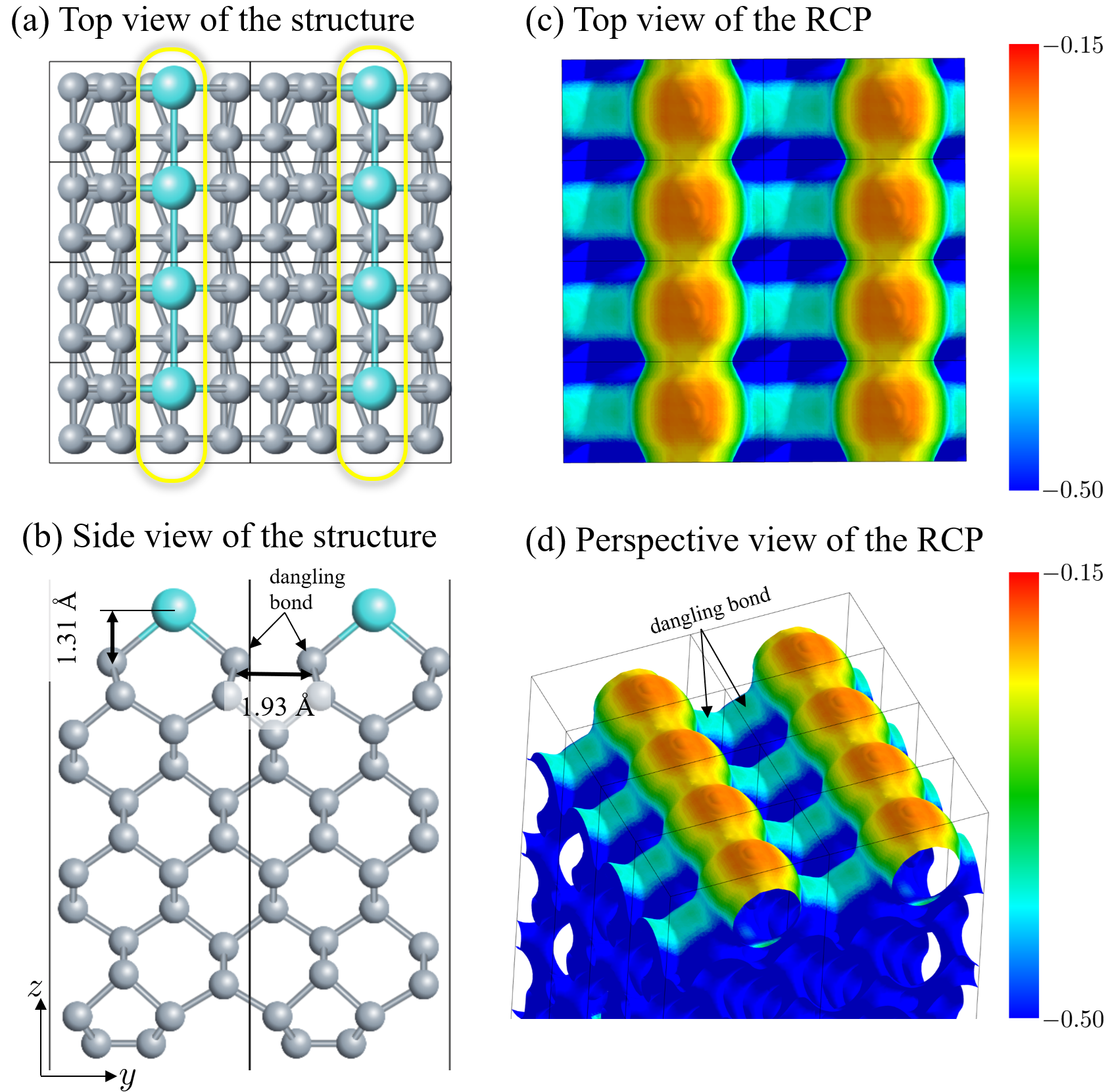
        }
        \caption{The optimized structure and the RCP analysis of the Si stripe structure on the diamond (001) surface.}
        \label{fig:Si_stripe_structure_RCP}
    \end{figure}

\begin{figure}[htbp]
        \centering
        \includegraphics[width=0.45\textwidth]{
            \PATHFIG/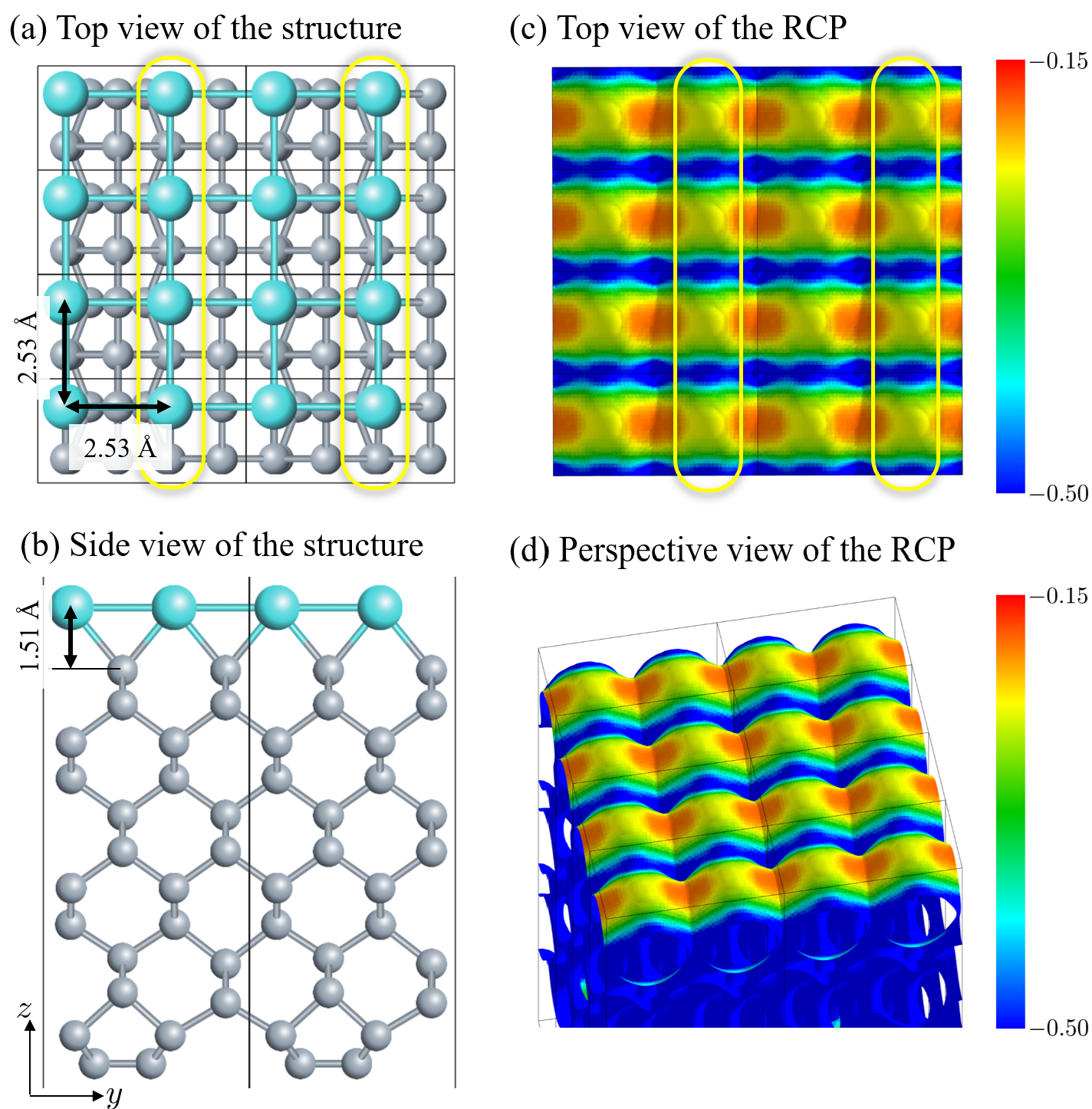
        }
        \caption{The optimized structure and the RCP analysis of the Si square lattice on the diamond (001) surface.}
        \label{fig:Si_square_lattice_RCP}
    \end{figure}

    To discuss electronic properties such as the band structure with higher accuracy, the thickness of the diamond slab was increased, and the bottom surface was terminated by carbon dimers.
    The optimized structures of the Si stripe structure and the Si square lattice, together with the corresponding RCP analysis, are shown in Figs.~\ref{fig:Si_stripe_structure_RCP} and \ref{fig:Si_square_lattice_RCP}. A high RCP is observed along the Si chains in the stripe structure as shown in Figs.~\ref{fig:Si_stripe_structure_RCP} (c) and (d).
    In addition, a weaker RCP distribution is found between adjacent chains, corresponding to bonding interactions between carbon atoms in the substrate.
    \rev{
    The C–C bond length of 1.94 \AA, as shown in Fig.~\ref{fig:Si_stripe_structure_RCP}(b), is longer than the typical C–C single-bond length of 1.55 \AA. This suggests an intermediate state in which the dangling bonds tend to bond with each other but have not yet formed a fully developed covalent bond.
    }
    This situation is analogous to that observed in the AA configuration of the carbon dimer ribbon as shown in Fig.~\ref{fig:RCP_C_ribbons} (c).
    The reason why the additional Si atoms occupy positions between the chains rather than directly above them during the formation of the Si square lattice is that such sites allow coordination with multiple neighboring atoms, including both Si and C atoms, thereby increasing the coordination number and stabilizing the structure.

    Although the surface Si atoms in the Si square lattice exhibit apparent fourfold symmetry, the underlying diamond substrate lacks fourfold rotational symmetry.
    Consequently, the RCP distribution does not possess fourfold symmetry.
    The RCP analysis shown in Figs.~\ref{fig:Si_square_lattice_RCP} (c) and (d) further demonstrates that Si atoms form bonds with both neighboring Si atoms and substrate carbon atoms, as evidenced by regions of high RCP (red) located near specific carbon atoms.
    The bonding characteristics between Si and C are also supported by the analysis of closest Wannier functions (CWFs)~\cite{PhysRevB.110.125115}.
    As shown in Figs.~S1 and S2 in Supplemental Materials (SM), the spatial distributions of the CWFs corresponding to the C-$s$, C-$p_y$, C-$p_z$, Si-$s$, Si-$p_y$, and Si-$p_z$ orbitals indicate significant hybridization between C and Si atoms.
    
    %%%%%%%%%%%%%%%%%%%%%%%%%%%%%%%%%%%%%%%%
    \subsection{Band analysis} \label{sec:band_analysis}
    %%%%%%%%%%%%%%%%%%%%%%%%%%%%%%%%%%%%%%%%
    
\begin{figure*}[htbp]
        \centering
        \includegraphics[width=0.95\textwidth]{
            \PATHFIG/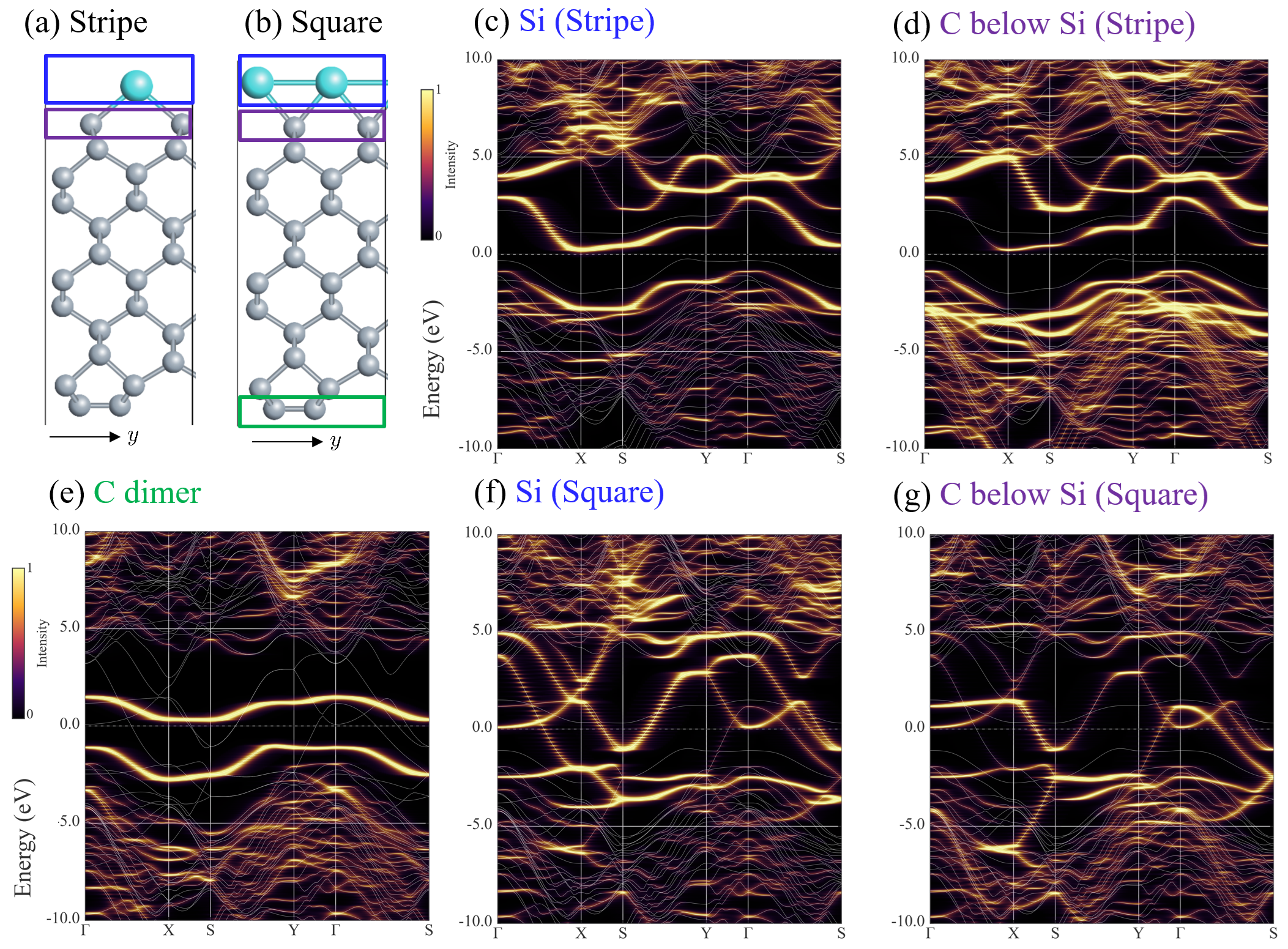
        }
        \caption{Band analysis of (a) the Si stripe structure and (b) the Si square lattice on the diamond (001) surface. 
        Atom-projected spectral weights are shown for the Si stripe structure in (c) and (d), projected onto the Si atom (blue square) and the C atom (purple square) in (a), respectively,
        and for the Si square lattice in (e)–(g), projected onto the C atom (green square), the Si atom (blue square), and the C atom (purple square) in (b), respectively.
        Thin white lines indicate the total band structure.
        }
        \label{fig:band_Si_C001}
    \end{figure*}

    Next, we investigate the electronic conduction characteristics using the band structures. Figure \ref{fig:band_Si_C001} shows calculated band structures and atomic-projected spectral weights of the Si stripe structure and the Si square lattice.
    The clean C-dimer surface is a semiconductor with a finite band gap. However, the atom-projected band analysis reveals two bands located near the Fermi level.
    Because the C dimer lies parallel to the substrate, the two carbon atoms contribute equally to these bands, which correspond to the bonding and antibonding states formed by the dimer. In contrast, 
    the Si-dimer-reconstructed Si(001) surface
    is known to exhibit dimer buckling, resulting in inequivalent contributions from the two Si atoms to different bands.
    Nevertheless, the Si-dimer surface also remains semiconducting with a finite band gap~\cite{PhysRevLett.74.1155,PhysRevB.54.13759} (see Fig.~S4 in SM).
    
    When the Si stripe structure is formed, the band gap is further reduced \rev{compared with that of the C-dimer surface}.
    For the Si square lattice, the band gap disappears completely
    with several bands crossing the Fermi level, indicating metallic behavior.
    The metallic nature of a freestanding Si square lattice monolayer has also been reported by DFT calculations in Ref.~\cite{NURHIDAYAH2026101363}.
    Furthermore, because the underlying diamond (001) substrate lacks fourfold rotational symmetry,
    anisotropic electronic conduction is expected along the $x$- and $y$-directions.
    
\begin{figure*}[htbp]
        \centering
        \includegraphics[width=0.85\textwidth]{
            \PATHFIG/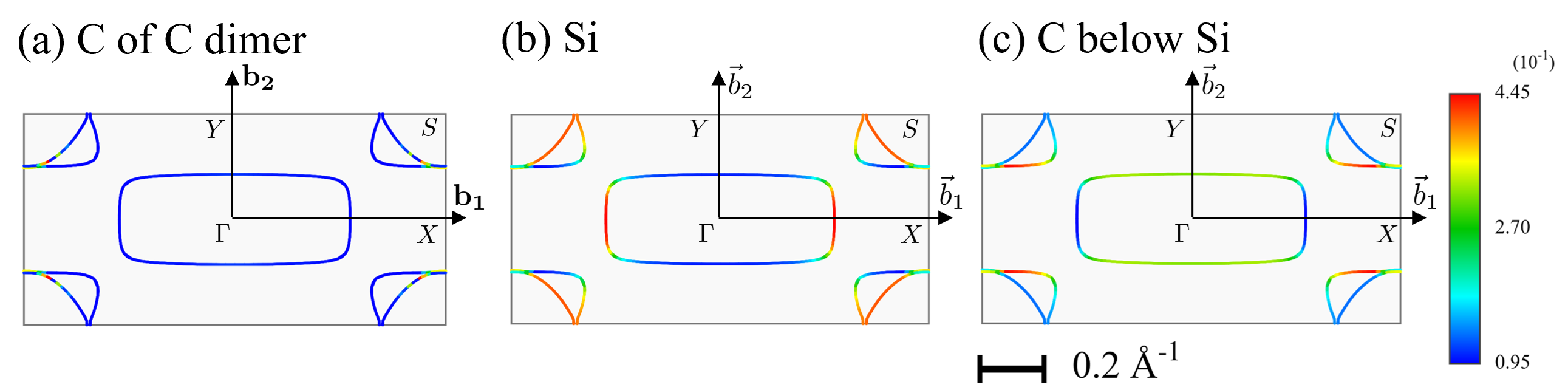
        }
        \caption{Fermi surface colored by atomic contributions.}
        \label{fig:FS_square_lattice}
    \end{figure*}

    Figure \ref{fig:FS_square_lattice} shows the Fermi surface colored by atomic contributions for the bands crossing the Fermi level.
    The states crossing the Fermi level along the $x$-direction have predominantly Si character, whereas those dispersing along the $y$-direction have substantial contributions from the substrate C atoms bonded to Si. These results demonstrate that the conducting states along the two in-plane directions originate from different atomic components.
    This interpretation is further supported by the orbital-projected band analysis shown in Fig.~S3 in SM.

    \subsection{Surface Phase Stability} \label{sec:phase_transition}
    %%%%%%%%%%%%%%%%%%%%%%%%%%%%%%%%%%%%%%%%

    Finally, we discuss the adsorption process of Si on the diamond (001) surface. In the case of low-temperature adsorption under a low Si flux, individual Si atoms are expected to adsorb sequentially, as discussed in Secs.~\ref{sec:Si_C001_energy} and \ref{sec:Si_C001_RCP}. As the Si coverage increases, the adsorbed atoms are expected to gradually form a stripe structure. With further increases in the Si supply, various intermediate structures may emerge before the system eventually transitions to the square-lattice structure.

\begin{figure}[htbp]
        \centering
        \includegraphics[width=0.45\textwidth]{
            \PATHFIG/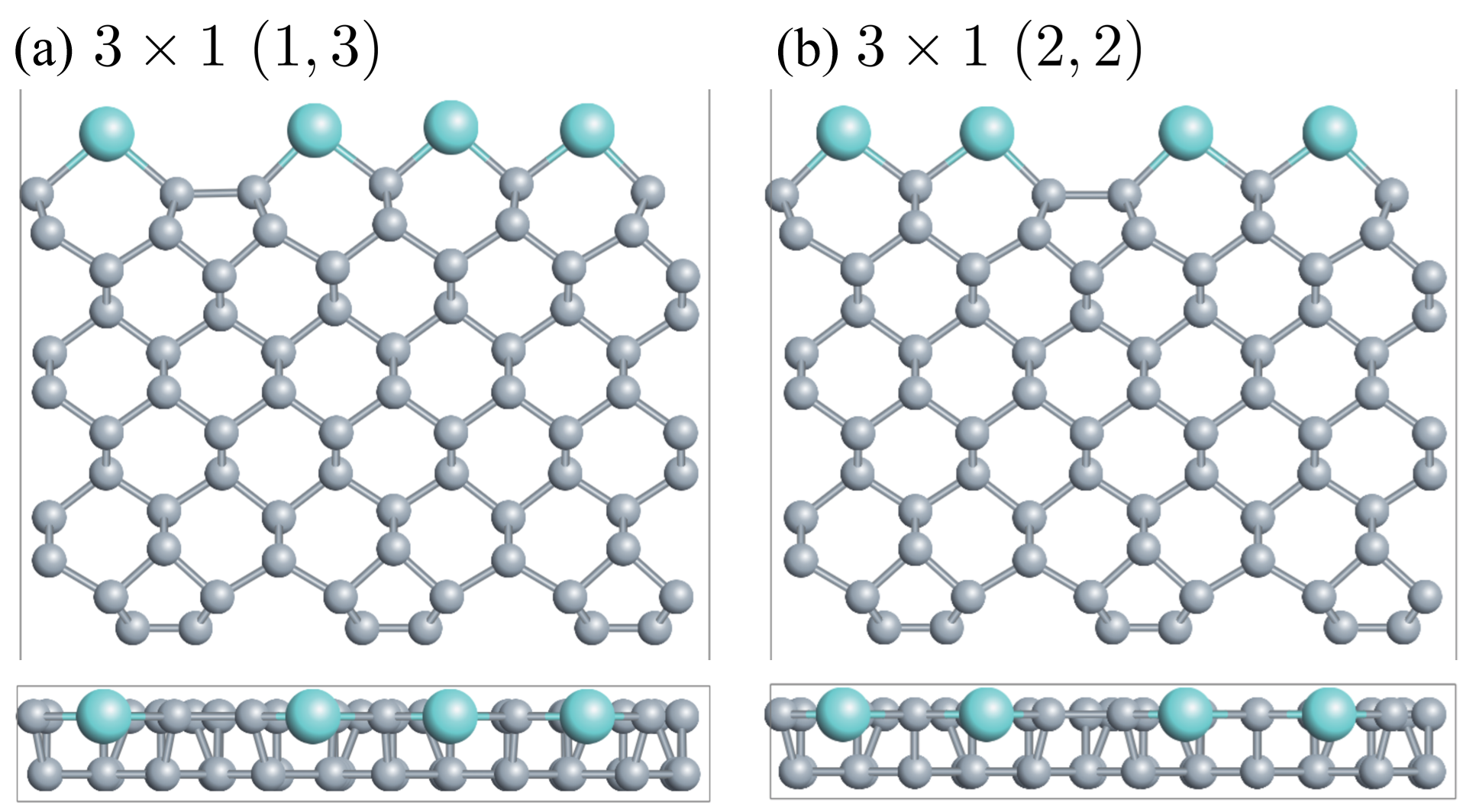
        }
        \caption{$3 \times 1(1,3)$ and $3 \times 1(2,2)$ structures.}
        \label{fig:3x1_structure}
    \end{figure}

    For example, when one additional Si wire is introduced into the $3 \times 1$ stripe structure, two possible configurations can be considered, as shown in Fig.~\ref{fig:3x1_structure}. The $3 \times 1\,(1,3)$ and $3 \times 1\,(2,2)$ structures correspond to arrangements in which Si wires are formed on the C-dimer surface with periodicities of $(1,3)$ and $(2,2)$, respectively. The adsorption energies obtained from geometry optimization calculations are summarized in Table~\ref{tab:energy_si_wires}. The $3 \times 1\,(2,2)$ structure is energetically more stable than the $3 \times 1\,(1,3)$ structure.
    Consistent with this energetic preference, the $3 \times 1\,(2,2)$ structure has been observed experimentally~\cite{10.1063/1.4921181,Schenk_2016} and investigated by first-principles calculations~\cite{10.1063/5.0203185}.
    %and has indeed been observed experimentally~\cite{10.1063/1.4921181,Schenk_2016},
    %and investigated by the first-principles calculations~\cite{10.1063/5.0203185}.

    However, since the $3 \times 1\,(2,2)$ arrangement cannot be obtained from the stripe structure simply by adding an Si chain without lateral rearrangement of the adsorbed Si atoms, the $3 \times 1\,(1,3)$ configuration is expected to form first.
    Subsequent thermally activated rearrangement of the Si atoms may then produce the energetically more stable $3 \times 1\,(2,2)$ structure.

\begin{figure}[htbp]
        \centering
        \includegraphics[width=0.48\textwidth]{
            \PATHFIG/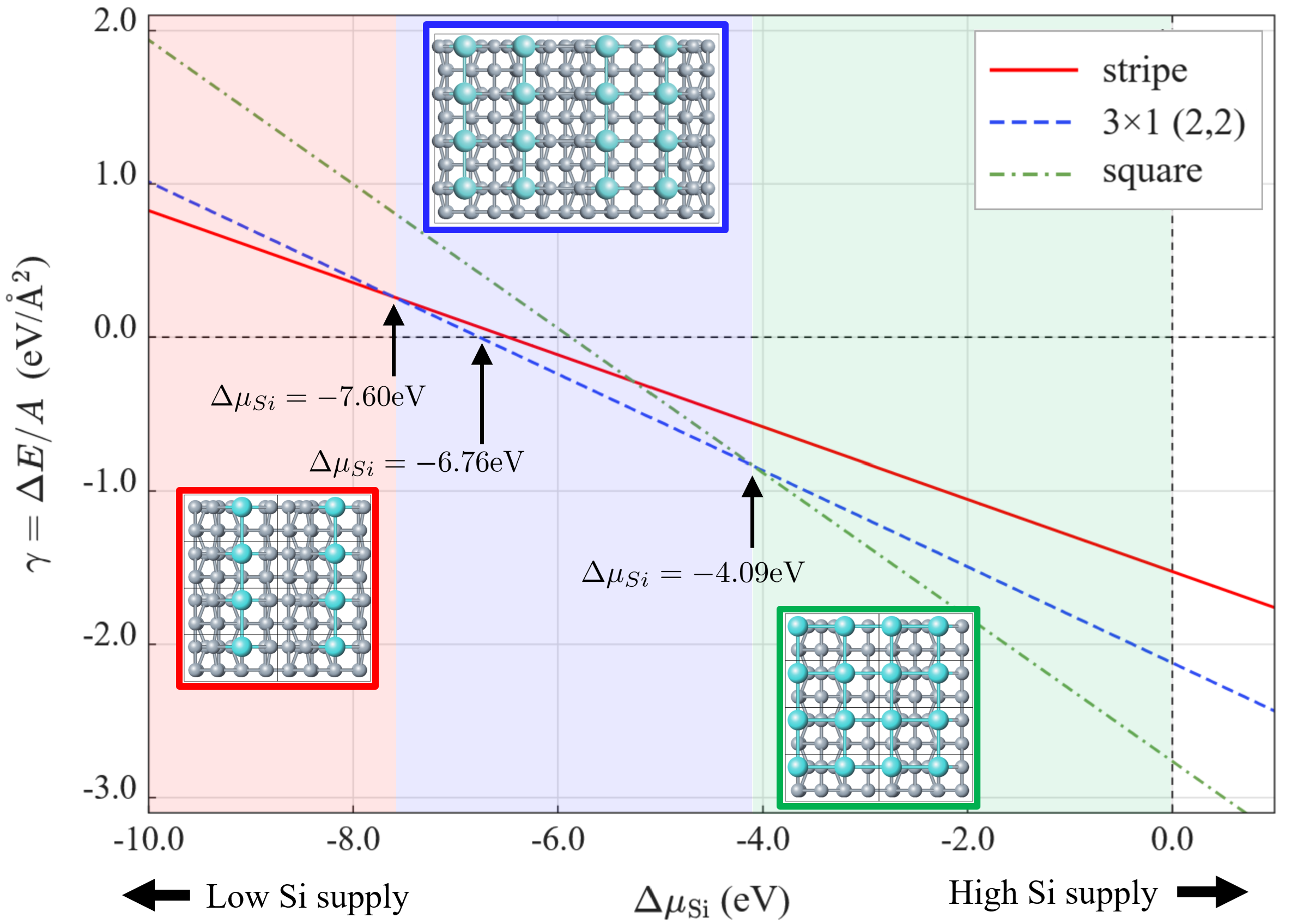
        }
        \caption{Surface phase stability $\gamma =\Delta E/A$ as a function of silicon chemical potential $\Delta \mu_{\rm Si}$.}
        \label{fig:phase_transition}
    \end{figure}

    Figure \ref{fig:phase_transition} shows the relative stability of the Si-covered surface structures as a function of the effective Si chemical potential.
    The surface formation energy is defined as
    $ \gamma=\Delta E / A =(E_{\rm ads}-n\Delta\mu_{\rm Si}) / A$,
    where
    $ \Delta\mu_{\rm Si} = \mu_{\rm Si}-E_{\rm Si\,atom}$,
    and $A$ is the surface area.
     Here, the Si chemical potential is referenced to the energy of an isolated Si atom, consistently with the definition of the adsorption energy used above. In the present analysis, $\Delta\mu_{\rm Si}$ is used as an effective parameter characterizing the Si-richness of the surface environment under Si supply, with increasing $\Delta\mu_{\rm Si}$ corresponding to increasingly Si-rich conditions.
    The results indicate that the $3\times1\,(2,2)$ structure becomes energetically favorable relative to the clean diamond surface when $\Delta\mu_{\rm Si}>-6.76$ eV. Furthermore, when $\Delta\mu_{\rm Si}>-4.09$ eV, the square-lattice structure becomes the most stable among the Si-covered structures considered. These results suggest a sequence of structural changes toward progressively higher Si coverages as the surface environment becomes increasingly Si rich.
    
    %Figure~\ref{fig:phase_transition} shows the surface phase stability as a function of the Si chemical potential. The surface formation energy is defined as
    %%$\gamma = \Delta E /A = \left(\left( E_{{\rm C(001)}+n{\rm Si}} - E_{\rm C(001)} - n E_{\rm Si \ atom} \right) - n \Delta \mu_{Si} \right)/A$,
    %$\gamma = \Delta E /A = \left( E_{\rm ads} - n \Delta \mu_{Si} \right)/A$,
    %which represents the thermodynamic stability of each surface structure under equilibrium conditions where Si atoms can be exchanged with an external reservoir.
    %Here, the Si chemical potential is referenced to a single Si atom rather than bulk Si because the experimentally relevant reservoir is an incident flux of atomic Si.
    %The results indicate that the $3 \times 1\,(2,2)$ structure becomes thermodynamically favorable when $\Delta\mu_{\mathrm{Si}} > -6.76~\mathrm{eV}$. Furthermore, when $\Delta\mu_{\mathrm{Si}} > -4.09~\mathrm{eV}$, the square-lattice structure becomes the most stable phase.
    %These results suggest a sequence of surface phase transitions driven by an increase in the Si chemical potential,
    %corresponding to increasingly Si-rich conditions associated with an increasing Si flux.

    \begin{table}[htbp]
      \centering
      \caption{Adsorption energies of the structures of Si wires on diamond (001) surface.}
      \label{tab:energy_si_wires}
      \begin{tabular}{lc}
        \hline
        System & $E_{\rm ads}$ (eV) \\
        \hline
        stripe structure & -19.46 \\
        $3\times1$ $(1,3)$ & -26.28 \\
        $3\times1$ $(2,2)$ & -27.07 \\
        square lattice & -35.25 \\
        \hline
      \end{tabular}
    \end{table}

    %%%%%%%%%%%%%%%%%%%%%%%%%%%%%%%%%%%%%%%%
    \section{Conclusions} \label{sec:conclusion}
    %%%%%%%%%%%%%%%%%%%%%%%%%%%%%%%%%%%%%%%%
    In this study, in addition to the energy-based analysis employed in previous studies,
    we have applied RCP analysis to the dimerized diamond (001) surface.
    We have demonstrated that RCP analysis can be used to elucidate surface bonding states and provide insight into crystal-growth mechanisms.
    Furthermore, the geometry optimizations have identified the preferential adsorption sites for Si adatoms on the diamond (001) surface.
    The adsorption of Si adatoms has been further investigated using RCP analysis,
    leading to the proposal that Si atoms are adsorbed in a stripe-like arrangement on the diamond (001) surface.
    With increasing Si coverage, the formation of a flat Si square-lattice structure has been predicted.
    
    A detailed band-structure analysis was performed
    for the predicted Si stripe structures and Si square lattice to investigate their electronic conduction characteristics.
    While the clean diamond (001) surface exhibits semiconducting behavior with a band gap,
    Si adsorption progressively reduces the surface band gap, which closes in the Si square lattice,
    resulting in metallic characteristics.
    Furthermore, the band analysis indicated that the Si square lattice exhibits anisotropic electronic conduction because the underlying diamond (001) substrate lacks fourfold rotational symmetry.
    %Although the thermodynamic stability has not been explicitly investigated in this study, the large adsorption energy of approximately 5 eV per Si atom indicates strong chemical bonding between Si and the C surface, suggesting that the proposed structures are energetically favorable and may be experimentally realizable.
    
    The Si adsorption process is expected to evolve from single adatoms to stripe structures with increasing Si coverage.
    Among the intermediate phases, the $3 \times 1\, (2,2)$ structure is energetically more stable than the $3 \times 1\, (1,3)$ structure,
    although thermally activated Si diffusion may be required for its formation.
    The surface phase analysis indicates that structures with progressively higher Si coverages become thermodynamically favored as the Si chemical potential increases, with the Si square-lattice structure eventually becoming the most stable among the structures considered.

    The present results demonstrate that RCP analysis can complement geometry optimization and total-energy comparison by identifying surface regions favorable for covalent-bond formation and thereby reducing the configurational search space for adsorption structures. This RCP-guided strategy may facilitate the prediction and interpretation of adsorption-driven structures on covalently bonded material surfaces.

%%%%%%%%%%%%%%%%%%%%%%%%%%%%%%%%%%%%%%%%%
\section{Acknowledgement}
%%%%%%%%%%%%%%%%%%%%%%%%%%%%%%%%%%%%%%%%%
The computation in this work has been done using the facilities of
the Supercomputer Center, the Institute for Solid State Physics, the
University of Tokyo (ISSPkyodo-SC-2025-Ba-0069, 2026-Ba-0086).

%%%%%%%%%%%%%%%%%%%%%%%%%%%%%%%%%%%%%%%%%
\section*{Author declarations}
%%%%%%%%%%%%%%%%%%%%%%%%%%%%%%%%%%%%%%%%%

\subsection*{Conflict of Interest}
The authors have no conflicts to disclose.

\subsection*{Author Contributions}

{\bf Masahiro Fukuda}:
Conceptualization (lead);
Funding acquisition (equal);
Data curation (lead);
Formal analysis (lead);
Investigation (equal);
Methodology (lead);
Resources (lead);
Software (lead);
Visualization (lead);
Writing - original draft preparation (equal);
Writing - review \& editing (equal).

{\bf Arath E. Marin Ramirez}:
Conceptualization (supporting);
Data curation (supporting);
Investigation (equal);
Writing - original draft preparation (equal);
Writing - review \& editing (equal).

{\bf Yoshiaki Sugimoto}:
Conceptualization (supporting);
Supervision (supporting);
Writing - review \& editing (equal).

{\bf Taisuke Ozaki}:
Conceptualization (supporting);
Funding acquisition (equal);
Software (supporting);
Supervision (equal);
Writing - review \& editing (equal).

%%%%%%%%%%%%%%%%%%%%%%%%%%%%%%%%%%%%%%%%%
\section*{Data availability statement}
%%%%%%%%%%%%%%%%%%%%%%%%%%%%%%%%%%%%%%%%%
The data that support the findings of this study are openly available in ISSP Data Repository at \url{https://isspns-gitlab.issp.u-tokyo.ac.jp/masahiro.fukuda/RCP_Analysis_of_Si_on_C001}.

    \bibliographystyle{apsrev4-2}
    \bibliography{ref}

\begin{thebibliography}{44}%
\makeatletter
\providecommand \@ifxundefined [1]{%
 \@ifx{#1\undefined}
}%
\providecommand \@ifnum [1]{%
 \ifnum #1\expandafter \@firstoftwo
 \else \expandafter \@secondoftwo
 \fi
}%
\providecommand \@ifx [1]{%
 \ifx #1\expandafter \@firstoftwo
 \else \expandafter \@secondoftwo
 \fi
}%
\providecommand \natexlab [1]{#1}%
\providecommand \enquote  [1]{``#1''}%
\providecommand \bibnamefont  [1]{#1}%
\providecommand \bibfnamefont [1]{#1}%
\providecommand \citenamefont [1]{#1}%
\providecommand \href@noop [0]{\@secondoftwo}%
\providecommand \href [0]{\begingroup \@sanitize@url \@href}%
\providecommand \@href[1]{\@@startlink{#1}\@@href}%
\providecommand \@@href[1]{\endgroup#1\@@endlink}%
\providecommand \@sanitize@url [0]{\catcode `\\12\catcode `\$12\catcode `\&12\catcode `\#12\catcode `\^12\catcode `\_12\catcode `\%12\relax}%
\providecommand \@@startlink[1]{}%
\providecommand \@@endlink[0]{}%
\providecommand \url  [0]{\begingroup\@sanitize@url \@url }%
\providecommand \@url [1]{\endgroup\@href {#1}{\urlprefix }}%
\providecommand \urlprefix  [0]{URL }%
\providecommand \Eprint [0]{\href }%
\providecommand \doibase [0]{https://doi.org/}%
\providecommand \selectlanguage [0]{\@gobble}%
\providecommand \bibinfo  [0]{\@secondoftwo}%
\providecommand \bibfield  [0]{\@secondoftwo}%
\providecommand \translation [1]{[#1]}%
\providecommand \BibitemOpen [0]{}%
\providecommand \bibitemStop [0]{}%
\providecommand \bibitemNoStop [0]{.\EOS\space}%
\providecommand \EOS [0]{\spacefactor3000\relax}%
\providecommand \BibitemShut  [1]{\csname bibitem#1\endcsname}%
\let\auto@bib@innerbib\@empty
%</preamble>
\bibitem [{\citenamefont {May}\ and\ \citenamefont {Zulkharnay}(2025)}]{10.1098/rsta.2023.0382}%
  \BibitemOpen
  \bibfield  {author} {\bibinfo {author} {\bibfnamefont {P.~W.}\ \bibnamefont {May}}\ and\ \bibinfo {author} {\bibfnamefont {R.}~\bibnamefont {Zulkharnay}},\ }\href {https://doi.org/10.1098/rsta.2023.0382} {\bibfield  {journal} {\bibinfo  {journal} {Philosophical Transactions of the Royal Society A: Mathematical, Physical and Engineering Sciences}\ }\textbf {\bibinfo {volume} {383}},\ \bibinfo {pages} {20230382} (\bibinfo {year} {2025})}\BibitemShut {NoStop}%
\bibitem [{\citenamefont {Ristein}(2006)}]{Ristein2006}%
  \BibitemOpen
  \bibfield  {author} {\bibinfo {author} {\bibfnamefont {J.}~\bibnamefont {Ristein}},\ }\href {https://doi.org/10.1007/s00339-005-3363-5} {\bibfield  {journal} {\bibinfo  {journal} {Applied Physics A}\ }\textbf {\bibinfo {volume} {82}},\ \bibinfo {pages} {377} (\bibinfo {year} {2006})}\BibitemShut {NoStop}%
\bibitem [{\citenamefont {Alfonso}\ \emph {et~al.}(1995)\citenamefont {Alfonso}, \citenamefont {Drabold},\ and\ \citenamefont {Ulloa}}]{PhysRevB.51.14669}%
  \BibitemOpen
  \bibfield  {author} {\bibinfo {author} {\bibfnamefont {D.~R.}\ \bibnamefont {Alfonso}}, \bibinfo {author} {\bibfnamefont {D.~A.}\ \bibnamefont {Drabold}},\ and\ \bibinfo {author} {\bibfnamefont {S.~E.}\ \bibnamefont {Ulloa}},\ }\href {https://doi.org/10.1103/PhysRevB.51.14669} {\bibfield  {journal} {\bibinfo  {journal} {Phys. Rev. B}\ }\textbf {\bibinfo {volume} {51}},\ \bibinfo {pages} {14669} (\bibinfo {year} {1995})}\BibitemShut {NoStop}%
\bibitem [{\citenamefont {Zhang}\ \emph {et~al.}(2025{\natexlab{a}})\citenamefont {Zhang}, \citenamefont {Yasui}, \citenamefont {Fukuda}, \citenamefont {Ozaki}, \citenamefont {Ogura}, \citenamefont {Makino}, \citenamefont {Takeuchi},\ and\ \citenamefont {Sugimoto}}]{Zhang2025}%
  \BibitemOpen
  \bibfield  {author} {\bibinfo {author} {\bibfnamefont {R.}~\bibnamefont {Zhang}}, \bibinfo {author} {\bibfnamefont {Y.}~\bibnamefont {Yasui}}, \bibinfo {author} {\bibfnamefont {M.}~\bibnamefont {Fukuda}}, \bibinfo {author} {\bibfnamefont {T.}~\bibnamefont {Ozaki}}, \bibinfo {author} {\bibfnamefont {M.}~\bibnamefont {Ogura}}, \bibinfo {author} {\bibfnamefont {T.}~\bibnamefont {Makino}}, \bibinfo {author} {\bibfnamefont {D.}~\bibnamefont {Takeuchi}},\ and\ \bibinfo {author} {\bibfnamefont {Y.}~\bibnamefont {Sugimoto}},\ }\href {https://doi.org/10.1021/acs.nanolett.4c05395} {\bibfield  {journal} {\bibinfo  {journal} {Nano Letters}\ }\textbf {\bibinfo {volume} {25}},\ \bibinfo {pages} {1101} (\bibinfo {year} {2025}{\natexlab{a}})}\BibitemShut {NoStop}%
\bibitem [{\citenamefont {Zhang}\ \emph {et~al.}(2025{\natexlab{b}})\citenamefont {Zhang}, \citenamefont {Yasui}, \citenamefont {Fukuda}, \citenamefont {Ogura}, \citenamefont {Makino}, \citenamefont {Takeuchi}, \citenamefont {Ozaki},\ and\ \citenamefont {Sugimoto}}]{PhysRevResearch.7.023036}%
  \BibitemOpen
  \bibfield  {author} {\bibinfo {author} {\bibfnamefont {R.}~\bibnamefont {Zhang}}, \bibinfo {author} {\bibfnamefont {Y.}~\bibnamefont {Yasui}}, \bibinfo {author} {\bibfnamefont {M.}~\bibnamefont {Fukuda}}, \bibinfo {author} {\bibfnamefont {M.}~\bibnamefont {Ogura}}, \bibinfo {author} {\bibfnamefont {T.}~\bibnamefont {Makino}}, \bibinfo {author} {\bibfnamefont {D.}~\bibnamefont {Takeuchi}}, \bibinfo {author} {\bibfnamefont {T.}~\bibnamefont {Ozaki}},\ and\ \bibinfo {author} {\bibfnamefont {Y.}~\bibnamefont {Sugimoto}},\ }\href {https://doi.org/10.1103/PhysRevResearch.7.023036} {\bibfield  {journal} {\bibinfo  {journal} {Phys. Rev. Res.}\ }\textbf {\bibinfo {volume} {7}},\ \bibinfo {pages} {023036} (\bibinfo {year} {2025}{\natexlab{b}})}\BibitemShut {NoStop}%
\bibitem [{\citenamefont {Fukuda}\ \emph {et~al.}(2026)\citenamefont {Fukuda}, \citenamefont {Senami}, \citenamefont {Sugimoto},\ and\ \citenamefont {Ozaki}}]{10.1063/5.0288934}%
  \BibitemOpen
  \bibfield  {author} {\bibinfo {author} {\bibfnamefont {M.}~\bibnamefont {Fukuda}}, \bibinfo {author} {\bibfnamefont {M.}~\bibnamefont {Senami}}, \bibinfo {author} {\bibfnamefont {Y.}~\bibnamefont {Sugimoto}},\ and\ \bibinfo {author} {\bibfnamefont {T.}~\bibnamefont {Ozaki}},\ }\href {https://doi.org/10.1063/5.0288934} {\bibfield  {journal} {\bibinfo  {journal} {The Journal of Chemical Physics}\ }\textbf {\bibinfo {volume} {164}},\ \bibinfo {pages} {084123} (\bibinfo {year} {2026})}\BibitemShut {NoStop}%
\bibitem [{\citenamefont {lin Chen}\ \emph {et~al.}(2012)\citenamefont {lin Chen}, \citenamefont {Shen}, \citenamefont {guo Zhang}, \citenamefont {Wang},\ and\ \citenamefont {hong Sun}}]{CHEN20123021}%
  \BibitemOpen
  \bibfield  {author} {\bibinfo {author} {\bibfnamefont {S.}~\bibnamefont {lin Chen}}, \bibinfo {author} {\bibfnamefont {B.}~\bibnamefont {Shen}}, \bibinfo {author} {\bibfnamefont {J.}~\bibnamefont {guo Zhang}}, \bibinfo {author} {\bibfnamefont {L.}~\bibnamefont {Wang}},\ and\ \bibinfo {author} {\bibfnamefont {F.}~\bibnamefont {hong Sun}},\ }\href {https://doi.org/https://doi.org/10.1016/S1003-6326(11)61565-3} {\bibfield  {journal} {\bibinfo  {journal} {Transactions of Nonferrous Metals Society of China}\ }\textbf {\bibinfo {volume} {22}},\ \bibinfo {pages} {3021} (\bibinfo {year} {2012})}\BibitemShut {NoStop}%
\bibitem [{\citenamefont {Yang}\ \emph {et~al.}(2022)\citenamefont {Yang}, \citenamefont {Ma},\ and\ \citenamefont {Dai}}]{yangProgressStructuralElectronic2022}%
  \BibitemOpen
  \bibfield  {author} {\bibinfo {author} {\bibfnamefont {H.}~\bibnamefont {Yang}}, \bibinfo {author} {\bibfnamefont {Y.}~\bibnamefont {Ma}},\ and\ \bibinfo {author} {\bibfnamefont {Y.}~\bibnamefont {Dai}},\ }\href {https://doi.org/10.1080/26941112.2021.1956287} {\bibfield  {journal} {\bibinfo  {journal} {Functional Diamond}\ }\textbf {\bibinfo {volume} {1}},\ \bibinfo {pages} {150} (\bibinfo {year} {2022})}\BibitemShut {NoStop}%
\bibitem [{\citenamefont {Edholm}\ \emph {et~al.}(1997)\citenamefont {Edholm}, \citenamefont {Olsson}, \citenamefont {Keskitalo},\ and\ \citenamefont {Vestling}}]{EDHOLM1997245}%
  \BibitemOpen
  \bibfield  {author} {\bibinfo {author} {\bibfnamefont {B.}~\bibnamefont {Edholm}}, \bibinfo {author} {\bibfnamefont {J.}~\bibnamefont {Olsson}}, \bibinfo {author} {\bibfnamefont {N.}~\bibnamefont {Keskitalo}},\ and\ \bibinfo {author} {\bibfnamefont {L.}~\bibnamefont {Vestling}},\ }\href {https://doi.org/https://doi.org/10.1016/S0167-9317(97)00057-9} {\bibfield  {journal} {\bibinfo  {journal} {Microelectronic Engineering}\ }\textbf {\bibinfo {volume} {36}},\ \bibinfo {pages} {245} (\bibinfo {year} {1997})},\ \bibinfo {note} {proceedings of the biennial conference on Insulating Films on Semiconductors}\BibitemShut {NoStop}%
\bibitem [{\citenamefont {Tsai}\ \emph {et~al.}(2020)\citenamefont {Tsai}, \citenamefont {Aghajamali}, \citenamefont {Dontschuk}, \citenamefont {Johnson}, \citenamefont {Usman}, \citenamefont {Schenk}, \citenamefont {Sear}, \citenamefont {Pakes}, \citenamefont {Hollenberg}, \citenamefont {McCallum}, \citenamefont {Rubanov}, \citenamefont {Tadich}, \citenamefont {Marks},\ and\ \citenamefont {Stacey}}]{10.1021/acsaelm.0c00289}%
  \BibitemOpen
  \bibfield  {author} {\bibinfo {author} {\bibfnamefont {A.}~\bibnamefont {Tsai}}, \bibinfo {author} {\bibfnamefont {A.}~\bibnamefont {Aghajamali}}, \bibinfo {author} {\bibfnamefont {N.}~\bibnamefont {Dontschuk}}, \bibinfo {author} {\bibfnamefont {B.~C.}\ \bibnamefont {Johnson}}, \bibinfo {author} {\bibfnamefont {M.}~\bibnamefont {Usman}}, \bibinfo {author} {\bibfnamefont {A.~K.}\ \bibnamefont {Schenk}}, \bibinfo {author} {\bibfnamefont {M.}~\bibnamefont {Sear}}, \bibinfo {author} {\bibfnamefont {C.~I.}\ \bibnamefont {Pakes}}, \bibinfo {author} {\bibfnamefont {L.~C.~L.}\ \bibnamefont {Hollenberg}}, \bibinfo {author} {\bibfnamefont {J.~C.}\ \bibnamefont {McCallum}}, \bibinfo {author} {\bibfnamefont {S.}~\bibnamefont {Rubanov}}, \bibinfo {author} {\bibfnamefont {A.}~\bibnamefont {Tadich}}, \bibinfo {author} {\bibfnamefont {N.~A.}\ \bibnamefont {Marks}},\ and\ \bibinfo {author} {\bibfnamefont {A.}~\bibnamefont {Stacey}},\ }\href {https://doi.org/10.1021/acsaelm.0c00289} {\bibfield  {journal} {\bibinfo  {journal}
  {ACS Applied Electronic Materials}\ }\textbf {\bibinfo {volume} {2}},\ \bibinfo {pages} {2003} (\bibinfo {year} {2020})}\BibitemShut {NoStop}%
\bibitem [{\citenamefont {Zhang}\ \emph {et~al.}(2025{\natexlab{c}})\citenamefont {Zhang}, \citenamefont {Peng},\ and\ \citenamefont {Ye}}]{10.1063/5.0270223}%
  \BibitemOpen
  \bibfield  {author} {\bibinfo {author} {\bibfnamefont {R.}~\bibnamefont {Zhang}}, \bibinfo {author} {\bibfnamefont {N.}~\bibnamefont {Peng}},\ and\ \bibinfo {author} {\bibfnamefont {H.}~\bibnamefont {Ye}},\ }\href {https://doi.org/10.1063/5.0270223} {\bibfield  {journal} {\bibinfo  {journal} {Applied Physics Letters}\ }\textbf {\bibinfo {volume} {127}},\ \bibinfo {pages} {070501} (\bibinfo {year} {2025}{\natexlab{c}})}\BibitemShut {NoStop}%
\bibitem [{\citenamefont {Cowie}\ \emph {et~al.}(2026)\citenamefont {Cowie}, \citenamefont {Deimert}, \citenamefont {Groome}, \citenamefont {Inayeh}, \citenamefont {Kirby}, \citenamefont {Mackie}, \citenamefont {Myall}, \citenamefont {Rohe}, \citenamefont {Sandoval}, \citenamefont {Sayed-Akhmad}, \citenamefont {Thanabalasingam}, \citenamefont {Wotton}, \citenamefont {Addou}, \citenamefont {Asani}, \citenamefont {Blue}, \citenamefont {Bottomley}, \citenamefont {Clarcia}, \citenamefont {Enright}, \citenamefont {Fan}, \citenamefont {Jr.}, \citenamefont {Godfrey}, \citenamefont {Guo}, \citenamefont {Hill}, \citenamefont {Huff}, \citenamefont {Jobes}, \citenamefont {Ma}, \citenamefont {Maahs}, \citenamefont {MacLean}, \citenamefont {Maley}, \citenamefont {Marshall}, \citenamefont {McCallum}, \citenamefont {Merkle}, \citenamefont {Morin}, \citenamefont {Plumadore}, \citenamefont {Rodriguez}, \citenamefont {Savoie}, \citenamefont {Scheffel}, \citenamefont {Wong}, \citenamefont {Allis}, \citenamefont {Barton},
  \citenamefont {Drew}, \citenamefont {Kennedy}, \citenamefont {Takatani}, \citenamefont {Taucer}, \citenamefont {Vobornik}, \citenamefont {Yamachika},\ and\ \citenamefont {Durand}}]{cowie2026atomicallyprecisemechanosynthesiscarbon}%
  \BibitemOpen
  \bibfield  {author} {\bibinfo {author} {\bibfnamefont {M.}~\bibnamefont {Cowie}}, \bibinfo {author} {\bibfnamefont {C.}~\bibnamefont {Deimert}}, \bibinfo {author} {\bibfnamefont {R.}~\bibnamefont {Groome}}, \bibinfo {author} {\bibfnamefont {A.}~\bibnamefont {Inayeh}}, \bibinfo {author} {\bibfnamefont {R.~J.}\ \bibnamefont {Kirby}}, \bibinfo {author} {\bibfnamefont {C.~J.}\ \bibnamefont {Mackie}}, \bibinfo {author} {\bibfnamefont {J.}~\bibnamefont {Myall}}, \bibinfo {author} {\bibfnamefont {S.}~\bibnamefont {Rohe}}, \bibinfo {author} {\bibfnamefont {L.}~\bibnamefont {Sandoval}}, \bibinfo {author} {\bibfnamefont {K.}~\bibnamefont {Sayed-Akhmad}}, \bibinfo {author} {\bibfnamefont {B.}~\bibnamefont {Thanabalasingam}}, \bibinfo {author} {\bibfnamefont {R.}~\bibnamefont {Wotton}}, \bibinfo {author} {\bibfnamefont {R.}~\bibnamefont {Addou}}, \bibinfo {author} {\bibfnamefont {A.}~\bibnamefont {Asani}}, \bibinfo {author} {\bibfnamefont {B.}~\bibnamefont {Blue}}, \bibinfo {author} {\bibfnamefont {A.}~\bibnamefont
  {Bottomley}}, \bibinfo {author} {\bibfnamefont {K.~A.}\ \bibnamefont {Clarcia}}, \bibinfo {author} {\bibfnamefont {T.}~\bibnamefont {Enright}}, \bibinfo {author} {\bibfnamefont {J.~Z.}\ \bibnamefont {Fan}}, \bibinfo {author} {\bibfnamefont {R.~A.~F.}\ \bibnamefont {Jr.}}, \bibinfo {author} {\bibfnamefont {A.~T.~K.}\ \bibnamefont {Godfrey}}, \bibinfo {author} {\bibfnamefont {S.~Y.}\ \bibnamefont {Guo}}, \bibinfo {author} {\bibfnamefont {A.}~\bibnamefont {Hill}}, \bibinfo {author} {\bibfnamefont {T.}~\bibnamefont {Huff}}, \bibinfo {author} {\bibfnamefont {M.}~\bibnamefont {Jobes}}, \bibinfo {author} {\bibfnamefont {H.}~\bibnamefont {Ma}}, \bibinfo {author} {\bibfnamefont {A.~C.}\ \bibnamefont {Maahs}}, \bibinfo {author} {\bibfnamefont {O.}~\bibnamefont {MacLean}}, \bibinfo {author} {\bibfnamefont {S.~M.}\ \bibnamefont {Maley}}, \bibinfo {author} {\bibfnamefont {M.}~\bibnamefont {Marshall}}, \bibinfo {author} {\bibfnamefont {T.}~\bibnamefont {McCallum}}, \bibinfo {author} {\bibfnamefont {R.~C.}\ \bibnamefont
  {Merkle}}, \bibinfo {author} {\bibfnamefont {M.}~\bibnamefont {Morin}}, \bibinfo {author} {\bibfnamefont {R.}~\bibnamefont {Plumadore}}, \bibinfo {author} {\bibfnamefont {H.}~\bibnamefont {Rodriguez}}, \bibinfo {author} {\bibfnamefont {M.}~\bibnamefont {Savoie}}, \bibinfo {author} {\bibfnamefont {B.}~\bibnamefont {Scheffel}}, \bibinfo {author} {\bibfnamefont {J.~L.}\ \bibnamefont {Wong}}, \bibinfo {author} {\bibfnamefont {D.~G.}\ \bibnamefont {Allis}}, \bibinfo {author} {\bibfnamefont {J.}~\bibnamefont {Barton}}, \bibinfo {author} {\bibfnamefont {M.}~\bibnamefont {Drew}}, \bibinfo {author} {\bibfnamefont {M.~R.}\ \bibnamefont {Kennedy}}, \bibinfo {author} {\bibfnamefont {T.}~\bibnamefont {Takatani}}, \bibinfo {author} {\bibfnamefont {M.}~\bibnamefont {Taucer}}, \bibinfo {author} {\bibfnamefont {D.}~\bibnamefont {Vobornik}}, \bibinfo {author} {\bibfnamefont {R.}~\bibnamefont {Yamachika}},\ and\ \bibinfo {author} {\bibfnamefont {M.}~\bibnamefont {Durand}},\ }\href {https://arxiv.org/abs/2605.27250} {\bibinfo
  {title} {Atomically precise mechanosynthesis of carbon structures on hydrogenated si(100) by inverted-mode stm}} (\bibinfo {year} {2026}),\ \Eprint {https://arxiv.org/abs/2605.27250} {arXiv:2605.27250 [cond-mat.mtrl-sci]} \BibitemShut {NoStop}%
\bibitem [{\citenamefont {Ren}\ \emph {et~al.}(2015)\citenamefont {Ren}, \citenamefont {Liu},\ and\ \citenamefont {Wei}}]{renEvolutionBehaviorSi2015}%
  \BibitemOpen
  \bibfield  {author} {\bibinfo {author} {\bibfnamefont {Y.}~\bibnamefont {Ren}}, \bibinfo {author} {\bibfnamefont {X.}~\bibnamefont {Liu}},\ and\ \bibinfo {author} {\bibfnamefont {H.}~\bibnamefont {Wei}},\ }\href {https://doi.org/10.1016/j.apsusc.2015.04.048} {\bibfield  {journal} {\bibinfo  {journal} {Applied Surface Science}\ }\textbf {\bibinfo {volume} {346}},\ \bibinfo {pages} {464} (\bibinfo {year} {2015})}\BibitemShut {NoStop}%
\bibitem [{\citenamefont {Kohn}\ and\ \citenamefont {Sham}(1965)}]{PhysRev.140.A1133}%
  \BibitemOpen
  \bibfield  {author} {\bibinfo {author} {\bibfnamefont {W.}~\bibnamefont {Kohn}}\ and\ \bibinfo {author} {\bibfnamefont {L.~J.}\ \bibnamefont {Sham}},\ }\href {https://doi.org/10.1103/PhysRev.140.A1133} {\bibfield  {journal} {\bibinfo  {journal} {Phys. Rev.}\ }\textbf {\bibinfo {volume} {140}},\ \bibinfo {pages} {A1133} (\bibinfo {year} {1965})}\BibitemShut {NoStop}%
\bibitem [{\citenamefont {Perdew}\ \emph {et~al.}(1996)\citenamefont {Perdew}, \citenamefont {Burke},\ and\ \citenamefont {Ernzerhof}}]{PhysRevLett.77.3865}%
  \BibitemOpen
  \bibfield  {author} {\bibinfo {author} {\bibfnamefont {J.~P.}\ \bibnamefont {Perdew}}, \bibinfo {author} {\bibfnamefont {K.}~\bibnamefont {Burke}},\ and\ \bibinfo {author} {\bibfnamefont {M.}~\bibnamefont {Ernzerhof}},\ }\href {https://doi.org/10.1103/PhysRevLett.77.3865} {\bibfield  {journal} {\bibinfo  {journal} {Phys. Rev. Lett.}\ }\textbf {\bibinfo {volume} {77}},\ \bibinfo {pages} {3865} (\bibinfo {year} {1996})}\BibitemShut {NoStop}%
\bibitem [{\citenamefont {Ozaki}\ \emph {et~al.}()\citenamefont {Ozaki} \emph {et~al.}}]{OpenMX}%
  \BibitemOpen
  \bibfield  {author} {\bibinfo {author} {\bibfnamefont {T.}~\bibnamefont {Ozaki}} \emph {et~al.},\ }\href@noop {} {\bibinfo {title} {\textsc{O}pen\textsc{MX} package}},\ \bibinfo {note} {\url{http://www.openmx-square.org/}}\BibitemShut {NoStop}%
\bibitem [{\citenamefont {Morrison}\ \emph {et~al.}(1993)\citenamefont {Morrison}, \citenamefont {Bylander},\ and\ \citenamefont {Kleinman}}]{PhysRevB.47.6728}%
  \BibitemOpen
  \bibfield  {author} {\bibinfo {author} {\bibfnamefont {I.}~\bibnamefont {Morrison}}, \bibinfo {author} {\bibfnamefont {D.~M.}\ \bibnamefont {Bylander}},\ and\ \bibinfo {author} {\bibfnamefont {L.}~\bibnamefont {Kleinman}},\ }\href {https://doi.org/10.1103/PhysRevB.47.6728} {\bibfield  {journal} {\bibinfo  {journal} {Phys. Rev. B}\ }\textbf {\bibinfo {volume} {47}},\ \bibinfo {pages} {6728} (\bibinfo {year} {1993})}\BibitemShut {NoStop}%
\bibitem [{\citenamefont {Ozaki}(2003)}]{PhysRevB.67.155108}%
  \BibitemOpen
  \bibfield  {author} {\bibinfo {author} {\bibfnamefont {T.}~\bibnamefont {Ozaki}},\ }\href {https://doi.org/10.1103/PhysRevB.67.155108} {\bibfield  {journal} {\bibinfo  {journal} {Phys. Rev. B}\ }\textbf {\bibinfo {volume} {67}},\ \bibinfo {pages} {155108} (\bibinfo {year} {2003})}\BibitemShut {NoStop}%
\bibitem [{\citenamefont {Lejaeghere}\ \emph {et~al.}(2016)\citenamefont {Lejaeghere}, \citenamefont {Bihlmayer}, \citenamefont {Bj{\"o}rkman}, \citenamefont {Blaha}, \citenamefont {Bl{\"u}gel}, \citenamefont {Blum}, \citenamefont {Caliste}, \citenamefont {Castelli}, \citenamefont {Clark}, \citenamefont {Dal~Corso}, \citenamefont {de~Gironcoli}, \citenamefont {Deutsch}, \citenamefont {Dewhurst}, \citenamefont {Di~Marco}, \citenamefont {Draxl}, \citenamefont {Du{\l}ak}, \citenamefont {Eriksson}, \citenamefont {Flores-Livas}, \citenamefont {Garrity}, \citenamefont {Genovese}, \citenamefont {Giannozzi}, \citenamefont {Giantomassi}, \citenamefont {Goedecker}, \citenamefont {Gonze}, \citenamefont {Gr{\aa}n{\"a}s}, \citenamefont {Gross}, \citenamefont {Gulans}, \citenamefont {Gygi}, \citenamefont {Hamann}, \citenamefont {Hasnip}, \citenamefont {Holzwarth}, \citenamefont {Iu{\c s}an}, \citenamefont {Jochym}, \citenamefont {Jollet}, \citenamefont {Jones}, \citenamefont {Kresse}, \citenamefont {Koepernik},
  \citenamefont {K{\"u}{\c c}{\"u}kbenli}, \citenamefont {Kvashnin}, \citenamefont {Locht}, \citenamefont {Lubeck}, \citenamefont {Marsman}, \citenamefont {Marzari}, \citenamefont {Nitzsche}, \citenamefont {Nordstr{\"o}m}, \citenamefont {Ozaki}, \citenamefont {Paulatto}, \citenamefont {Pickard}, \citenamefont {Poelmans}, \citenamefont {Probert}, \citenamefont {Refson}, \citenamefont {Richter}, \citenamefont {Rignanese}, \citenamefont {Saha}, \citenamefont {Scheffler}, \citenamefont {Schlipf}, \citenamefont {Schwarz}, \citenamefont {Sharma}, \citenamefont {Tavazza}, \citenamefont {Thunstr{\"o}m}, \citenamefont {Tkatchenko}, \citenamefont {Torrent}, \citenamefont {Vanderbilt}, \citenamefont {van Setten}, \citenamefont {Van~Speybroeck}, \citenamefont {Wills}, \citenamefont {Yates}, \citenamefont {Zhang},\ and\ \citenamefont {Cottenier}}]{Lejaeghere2016-gh}%
  \BibitemOpen
  \bibfield  {author} {\bibinfo {author} {\bibfnamefont {K.}~\bibnamefont {Lejaeghere}}, \bibinfo {author} {\bibfnamefont {G.}~\bibnamefont {Bihlmayer}}, \bibinfo {author} {\bibfnamefont {T.}~\bibnamefont {Bj{\"o}rkman}}, \bibinfo {author} {\bibfnamefont {P.}~\bibnamefont {Blaha}}, \bibinfo {author} {\bibfnamefont {S.}~\bibnamefont {Bl{\"u}gel}}, \bibinfo {author} {\bibfnamefont {V.}~\bibnamefont {Blum}}, \bibinfo {author} {\bibfnamefont {D.}~\bibnamefont {Caliste}}, \bibinfo {author} {\bibfnamefont {I.~E.}\ \bibnamefont {Castelli}}, \bibinfo {author} {\bibfnamefont {S.~J.}\ \bibnamefont {Clark}}, \bibinfo {author} {\bibfnamefont {A.}~\bibnamefont {Dal~Corso}}, \bibinfo {author} {\bibfnamefont {S.}~\bibnamefont {de~Gironcoli}}, \bibinfo {author} {\bibfnamefont {T.}~\bibnamefont {Deutsch}}, \bibinfo {author} {\bibfnamefont {J.~K.}\ \bibnamefont {Dewhurst}}, \bibinfo {author} {\bibfnamefont {I.}~\bibnamefont {Di~Marco}}, \bibinfo {author} {\bibfnamefont {C.}~\bibnamefont {Draxl}}, \bibinfo {author}
  {\bibfnamefont {M.}~\bibnamefont {Du{\l}ak}}, \bibinfo {author} {\bibfnamefont {O.}~\bibnamefont {Eriksson}}, \bibinfo {author} {\bibfnamefont {J.~A.}\ \bibnamefont {Flores-Livas}}, \bibinfo {author} {\bibfnamefont {K.~F.}\ \bibnamefont {Garrity}}, \bibinfo {author} {\bibfnamefont {L.}~\bibnamefont {Genovese}}, \bibinfo {author} {\bibfnamefont {P.}~\bibnamefont {Giannozzi}}, \bibinfo {author} {\bibfnamefont {M.}~\bibnamefont {Giantomassi}}, \bibinfo {author} {\bibfnamefont {S.}~\bibnamefont {Goedecker}}, \bibinfo {author} {\bibfnamefont {X.}~\bibnamefont {Gonze}}, \bibinfo {author} {\bibfnamefont {O.}~\bibnamefont {Gr{\aa}n{\"a}s}}, \bibinfo {author} {\bibfnamefont {E.~K.~U.}\ \bibnamefont {Gross}}, \bibinfo {author} {\bibfnamefont {A.}~\bibnamefont {Gulans}}, \bibinfo {author} {\bibfnamefont {F.}~\bibnamefont {Gygi}}, \bibinfo {author} {\bibfnamefont {D.~R.}\ \bibnamefont {Hamann}}, \bibinfo {author} {\bibfnamefont {P.~J.}\ \bibnamefont {Hasnip}}, \bibinfo {author} {\bibfnamefont {N.~A.~W.}\ \bibnamefont
  {Holzwarth}}, \bibinfo {author} {\bibfnamefont {D.}~\bibnamefont {Iu{\c s}an}}, \bibinfo {author} {\bibfnamefont {D.~B.}\ \bibnamefont {Jochym}}, \bibinfo {author} {\bibfnamefont {F.}~\bibnamefont {Jollet}}, \bibinfo {author} {\bibfnamefont {D.}~\bibnamefont {Jones}}, \bibinfo {author} {\bibfnamefont {G.}~\bibnamefont {Kresse}}, \bibinfo {author} {\bibfnamefont {K.}~\bibnamefont {Koepernik}}, \bibinfo {author} {\bibfnamefont {E.}~\bibnamefont {K{\"u}{\c c}{\"u}kbenli}}, \bibinfo {author} {\bibfnamefont {Y.~O.}\ \bibnamefont {Kvashnin}}, \bibinfo {author} {\bibfnamefont {I.~L.~M.}\ \bibnamefont {Locht}}, \bibinfo {author} {\bibfnamefont {S.}~\bibnamefont {Lubeck}}, \bibinfo {author} {\bibfnamefont {M.}~\bibnamefont {Marsman}}, \bibinfo {author} {\bibfnamefont {N.}~\bibnamefont {Marzari}}, \bibinfo {author} {\bibfnamefont {U.}~\bibnamefont {Nitzsche}}, \bibinfo {author} {\bibfnamefont {L.}~\bibnamefont {Nordstr{\"o}m}}, \bibinfo {author} {\bibfnamefont {T.}~\bibnamefont {Ozaki}}, \bibinfo {author}
  {\bibfnamefont {L.}~\bibnamefont {Paulatto}}, \bibinfo {author} {\bibfnamefont {C.~J.}\ \bibnamefont {Pickard}}, \bibinfo {author} {\bibfnamefont {W.}~\bibnamefont {Poelmans}}, \bibinfo {author} {\bibfnamefont {M.~I.~J.}\ \bibnamefont {Probert}}, \bibinfo {author} {\bibfnamefont {K.}~\bibnamefont {Refson}}, \bibinfo {author} {\bibfnamefont {M.}~\bibnamefont {Richter}}, \bibinfo {author} {\bibfnamefont {G.-M.}\ \bibnamefont {Rignanese}}, \bibinfo {author} {\bibfnamefont {S.}~\bibnamefont {Saha}}, \bibinfo {author} {\bibfnamefont {M.}~\bibnamefont {Scheffler}}, \bibinfo {author} {\bibfnamefont {M.}~\bibnamefont {Schlipf}}, \bibinfo {author} {\bibfnamefont {K.}~\bibnamefont {Schwarz}}, \bibinfo {author} {\bibfnamefont {S.}~\bibnamefont {Sharma}}, \bibinfo {author} {\bibfnamefont {F.}~\bibnamefont {Tavazza}}, \bibinfo {author} {\bibfnamefont {P.}~\bibnamefont {Thunstr{\"o}m}}, \bibinfo {author} {\bibfnamefont {A.}~\bibnamefont {Tkatchenko}}, \bibinfo {author} {\bibfnamefont {M.}~\bibnamefont {Torrent}},
  \bibinfo {author} {\bibfnamefont {D.}~\bibnamefont {Vanderbilt}}, \bibinfo {author} {\bibfnamefont {M.~J.}\ \bibnamefont {van Setten}}, \bibinfo {author} {\bibfnamefont {V.}~\bibnamefont {Van~Speybroeck}}, \bibinfo {author} {\bibfnamefont {J.~M.}\ \bibnamefont {Wills}}, \bibinfo {author} {\bibfnamefont {J.~R.}\ \bibnamefont {Yates}}, \bibinfo {author} {\bibfnamefont {G.-X.}\ \bibnamefont {Zhang}},\ and\ \bibinfo {author} {\bibfnamefont {S.}~\bibnamefont {Cottenier}},\ }\href@noop {} {\bibfield  {journal} {\bibinfo  {journal} {Science}\ }\textbf {\bibinfo {volume} {351}},\ \bibinfo {pages} {aad3000} (\bibinfo {year} {2016})}\BibitemShut {NoStop}%
\bibitem [{\citenamefont {Ozaki}\ and\ \citenamefont {Kino}(2005)}]{PhysRevB.72.045121}%
  \BibitemOpen
  \bibfield  {author} {\bibinfo {author} {\bibfnamefont {T.}~\bibnamefont {Ozaki}}\ and\ \bibinfo {author} {\bibfnamefont {H.}~\bibnamefont {Kino}},\ }\href {https://doi.org/10.1103/PhysRevB.72.045121} {\bibfield  {journal} {\bibinfo  {journal} {Phys. Rev. B}\ }\textbf {\bibinfo {volume} {72}},\ \bibinfo {pages} {045121} (\bibinfo {year} {2005})}\BibitemShut {NoStop}%
\bibitem [{\citenamefont {Banerjee}\ \emph {et~al.}(1985)\citenamefont {Banerjee}, \citenamefont {Adams}, \citenamefont {Simons},\ and\ \citenamefont {Shepard}}]{doi:10.1021/j100247a015}%
  \BibitemOpen
  \bibfield  {author} {\bibinfo {author} {\bibfnamefont {A.}~\bibnamefont {Banerjee}}, \bibinfo {author} {\bibfnamefont {N.}~\bibnamefont {Adams}}, \bibinfo {author} {\bibfnamefont {J.}~\bibnamefont {Simons}},\ and\ \bibinfo {author} {\bibfnamefont {R.}~\bibnamefont {Shepard}},\ }\href {https://doi.org/10.1021/j100247a015} {\bibfield  {journal} {\bibinfo  {journal} {The Journal of Physical Chemistry}\ }\textbf {\bibinfo {volume} {89}},\ \bibinfo {pages} {52} (\bibinfo {year} {1985})}\BibitemShut {NoStop}%
\bibitem [{\citenamefont {Csaszar}\ and\ \citenamefont {Pulay}(1984)}]{CSASZAR198431}%
  \BibitemOpen
  \bibfield  {author} {\bibinfo {author} {\bibfnamefont {P.}~\bibnamefont {Csaszar}}\ and\ \bibinfo {author} {\bibfnamefont {P.}~\bibnamefont {Pulay}},\ }\href {https://doi.org/https://doi.org/10.1016/S0022-2860(84)87198-7} {\bibfield  {journal} {\bibinfo  {journal} {Journal of Molecular Structure}\ }\textbf {\bibinfo {volume} {114}},\ \bibinfo {pages} {31} (\bibinfo {year} {1984})}\BibitemShut {NoStop}%
\bibitem [{\citenamefont {BROYDEN}(1970)}]{10.1093/imamat/6.1.76}%
  \BibitemOpen
  \bibfield  {author} {\bibinfo {author} {\bibfnamefont {C.~G.}\ \bibnamefont {BROYDEN}},\ }\href {https://doi.org/10.1093/imamat/6.1.76} {\bibfield  {journal} {\bibinfo  {journal} {IMA Journal of Applied Mathematics}\ }\textbf {\bibinfo {volume} {6}},\ \bibinfo {pages} {76} (\bibinfo {year} {1970})}\BibitemShut {NoStop}%
\bibitem [{\citenamefont {Fletcher}(1970)}]{10.1093/comjnl/13.3.317}%
  \BibitemOpen
  \bibfield  {author} {\bibinfo {author} {\bibfnamefont {R.}~\bibnamefont {Fletcher}},\ }\href {https://doi.org/10.1093/comjnl/13.3.317} {\bibfield  {journal} {\bibinfo  {journal} {The Computer Journal}\ }\textbf {\bibinfo {volume} {13}},\ \bibinfo {pages} {317} (\bibinfo {year} {1970})}\BibitemShut {NoStop}%
\bibitem [{\citenamefont {Goldfarb}(1970)}]{10.2307/2004873}%
  \BibitemOpen
  \bibfield  {author} {\bibinfo {author} {\bibfnamefont {D.}~\bibnamefont {Goldfarb}},\ }\href {http://www.jstor.org/stable/2004873} {\bibfield  {journal} {\bibinfo  {journal} {Mathematics of Computation}\ }\textbf {\bibinfo {volume} {24}},\ \bibinfo {pages} {23} (\bibinfo {year} {1970})}\BibitemShut {NoStop}%
\bibitem [{\citenamefont {Shanno}(1970)}]{10.2307/2004840}%
  \BibitemOpen
  \bibfield  {author} {\bibinfo {author} {\bibfnamefont {D.~F.}\ \bibnamefont {Shanno}},\ }\href {http://www.jstor.org/stable/2004840} {\bibfield  {journal} {\bibinfo  {journal} {Mathematics of Computation}\ }\textbf {\bibinfo {volume} {24}},\ \bibinfo {pages} {647} (\bibinfo {year} {1970})}\BibitemShut {NoStop}%
\bibitem [{\citenamefont {Fukuda}(2025{\natexlab{a}})}]{FLPQ}%
  \BibitemOpen
  \bibfield  {author} {\bibinfo {author} {\bibfnamefont {M.}~\bibnamefont {Fukuda}},\ }\href@noop {} {\bibinfo {title} {\textsc{FLPQ} package}} (\bibinfo {year} {2025}{\natexlab{a}}),\ \bibinfo {note} {\url{https://github.com/mfukudaQED/FLPQ}}\BibitemShut {NoStop}%
\bibitem [{\citenamefont {Senami}\ \emph {et~al.}()\citenamefont {Senami}, \citenamefont {Ichikawa}, \citenamefont {Tachibana}, \citenamefont {Fukuda},\ and\ \citenamefont {Soga}}]{QEDalpha}%
  \BibitemOpen
  \bibfield  {author} {\bibinfo {author} {\bibfnamefont {M.}~\bibnamefont {Senami}}, \bibinfo {author} {\bibfnamefont {K.}~\bibnamefont {Ichikawa}}, \bibinfo {author} {\bibfnamefont {A.}~\bibnamefont {Tachibana}}, \bibinfo {author} {\bibfnamefont {M.}~\bibnamefont {Fukuda}},\ and\ \bibinfo {author} {\bibfnamefont {K.}~\bibnamefont {Soga}},\ }\href@noop {} {\bibinfo {title} {\textsc{QEDalpha} package}},\ \bibinfo {note} {\url{https://github.com/mfukudaQED/QEDalpha}}\BibitemShut {NoStop}%
\bibitem [{\citenamefont {Ozaki}(2024)}]{PhysRevB.110.125115}%
  \BibitemOpen
  \bibfield  {author} {\bibinfo {author} {\bibfnamefont {T.}~\bibnamefont {Ozaki}},\ }\href {https://doi.org/10.1103/PhysRevB.110.125115} {\bibfield  {journal} {\bibinfo  {journal} {Phys. Rev. B}\ }\textbf {\bibinfo {volume} {110}},\ \bibinfo {pages} {125115} (\bibinfo {year} {2024})}\BibitemShut {NoStop}%
\bibitem [{\citenamefont {Lee}\ and\ \citenamefont {Ozaki}(2019)}]{LEE2019192}%
  \BibitemOpen
  \bibfield  {author} {\bibinfo {author} {\bibfnamefont {Y.-T.}\ \bibnamefont {Lee}}\ and\ \bibinfo {author} {\bibfnamefont {T.}~\bibnamefont {Ozaki}},\ }\href {https://doi.org/https://doi.org/10.1016/j.jmgm.2019.03.013} {\bibfield  {journal} {\bibinfo  {journal} {Journal of Molecular Graphics and Modelling}\ }\textbf {\bibinfo {volume} {89}},\ \bibinfo {pages} {192} (\bibinfo {year} {2019})}\BibitemShut {NoStop}%
\bibitem [{\citenamefont {Fukuda}(2025{\natexlab{b}})}]{FLPQViewer}%
  \BibitemOpen
  \bibfield  {author} {\bibinfo {author} {\bibfnamefont {M.}~\bibnamefont {Fukuda}},\ }\href@noop {} {\bibinfo {title} {\textsc{FLPQViewer}}} (\bibinfo {year} {2025}{\natexlab{b}}),\ \bibinfo {note} {\url{https://github.com/mfukudaQED/FLPQViewer}}\BibitemShut {NoStop}%
\bibitem [{\citenamefont {Fukuda}(2026)}]{PQViewer}%
  \BibitemOpen
  \bibfield  {author} {\bibinfo {author} {\bibfnamefont {M.}~\bibnamefont {Fukuda}},\ }\href@noop {} {\bibinfo {title} {\textsc{PQViewer}}} (\bibinfo {year} {2026}),\ \bibinfo {note} {\url{https://github.com/mfukudaQED/PQViewer}}\BibitemShut {NoStop}%
\bibitem [{\citenamefont {Tachibana}(2001)}]{Tachibana2001}%
  \BibitemOpen
  \bibfield  {author} {\bibinfo {author} {\bibfnamefont {A.}~\bibnamefont {Tachibana}},\ }\href {https://doi.org/10.1063/1.1384012} {\bibfield  {journal} {\bibinfo  {journal} {The Journal of Chemical Physics}\ }\textbf {\bibinfo {volume} {115}},\ \bibinfo {pages} {3497} (\bibinfo {year} {2001})}\BibitemShut {NoStop}%
\bibitem [{\citenamefont {Tachibana}\ and\ \citenamefont {Parr}(1992)}]{Tachibana1992}%
  \BibitemOpen
  \bibfield  {author} {\bibinfo {author} {\bibfnamefont {A.}~\bibnamefont {Tachibana}}\ and\ \bibinfo {author} {\bibfnamefont {R.~G.}\ \bibnamefont {Parr}},\ }\href {https://doi.org/https://doi.org/10.1002/qua.560410402} {\bibfield  {journal} {\bibinfo  {journal} {International Journal of Quantum Chemistry}\ }\textbf {\bibinfo {volume} {41}},\ \bibinfo {pages} {527} (\bibinfo {year} {1992})}\BibitemShut {NoStop}%
\bibitem [{\citenamefont {Tachibana}\ \emph {et~al.}(1999)\citenamefont {Tachibana}, \citenamefont {Nakamura}, \citenamefont {Sakata},\ and\ \citenamefont {Morisaki}}]{tachibana1999}%
  \BibitemOpen
  \bibfield  {author} {\bibinfo {author} {\bibfnamefont {A.}~\bibnamefont {Tachibana}}, \bibinfo {author} {\bibfnamefont {K.}~\bibnamefont {Nakamura}}, \bibinfo {author} {\bibfnamefont {K.}~\bibnamefont {Sakata}},\ and\ \bibinfo {author} {\bibfnamefont {T.}~\bibnamefont {Morisaki}},\ }\href {https://doi.org/https://doi.org/10.1002/(SICI)1097-461X(1999)74:6<669::AID-QUA8>3.0.CO;2-O} {\bibfield  {journal} {\bibinfo  {journal} {International Journal of Quantum Chemistry}\ }\textbf {\bibinfo {volume} {74}},\ \bibinfo {pages} {669} (\bibinfo {year} {1999})}\BibitemShut {NoStop}%
\bibitem [{\citenamefont {Szarek}\ \emph {et~al.}(2009)\citenamefont {Szarek}, \citenamefont {Urakami}, \citenamefont {Zhou}, \citenamefont {Cheng},\ and\ \citenamefont {Tachibana}}]{10.1063/1.3072369}%
  \BibitemOpen
  \bibfield  {author} {\bibinfo {author} {\bibfnamefont {P.}~\bibnamefont {Szarek}}, \bibinfo {author} {\bibfnamefont {K.}~\bibnamefont {Urakami}}, \bibinfo {author} {\bibfnamefont {C.}~\bibnamefont {Zhou}}, \bibinfo {author} {\bibfnamefont {H.}~\bibnamefont {Cheng}},\ and\ \bibinfo {author} {\bibfnamefont {A.}~\bibnamefont {Tachibana}},\ }\href {https://doi.org/10.1063/1.3072369} {\bibfield  {journal} {\bibinfo  {journal} {The Journal of Chemical Physics}\ }\textbf {\bibinfo {volume} {130}},\ \bibinfo {pages} {084111} (\bibinfo {year} {2009})}\BibitemShut {NoStop}%
\bibitem [{\citenamefont {Senami}\ \emph {et~al.}(2011)\citenamefont {Senami}, \citenamefont {Ikeda}, \citenamefont {Fukushima},\ and\ \citenamefont {Tachibana}}]{10.1063/1.3651182}%
  \BibitemOpen
  \bibfield  {author} {\bibinfo {author} {\bibfnamefont {M.}~\bibnamefont {Senami}}, \bibinfo {author} {\bibfnamefont {Y.}~\bibnamefont {Ikeda}}, \bibinfo {author} {\bibfnamefont {A.}~\bibnamefont {Fukushima}},\ and\ \bibinfo {author} {\bibfnamefont {A.}~\bibnamefont {Tachibana}},\ }\href {https://doi.org/10.1063/1.3651182} {\bibfield  {journal} {\bibinfo  {journal} {AIP Advances}\ }\textbf {\bibinfo {volume} {1}},\ \bibinfo {pages} {042106} (\bibinfo {year} {2011})}\BibitemShut {NoStop}%
\bibitem [{\citenamefont {Szarek}\ \emph {et~al.}(2008)\citenamefont {Szarek}, \citenamefont {Sueda},\ and\ \citenamefont {Tachibana}}]{10.1063/1.2973634}%
  \BibitemOpen
  \bibfield  {author} {\bibinfo {author} {\bibfnamefont {P.}~\bibnamefont {Szarek}}, \bibinfo {author} {\bibfnamefont {Y.}~\bibnamefont {Sueda}},\ and\ \bibinfo {author} {\bibfnamefont {A.}~\bibnamefont {Tachibana}},\ }\href {https://doi.org/10.1063/1.2973634} {\bibfield  {journal} {\bibinfo  {journal} {The Journal of Chemical Physics}\ }\textbf {\bibinfo {volume} {129}},\ \bibinfo {pages} {094102} (\bibinfo {year} {2008})}\BibitemShut {NoStop}%
\bibitem [{\citenamefont {Kr\"uger}\ and\ \citenamefont {Pollmann}(1995)}]{PhysRevLett.74.1155}%
  \BibitemOpen
  \bibfield  {author} {\bibinfo {author} {\bibfnamefont {P.}~\bibnamefont {Kr\"uger}}\ and\ \bibinfo {author} {\bibfnamefont {J.}~\bibnamefont {Pollmann}},\ }\href {https://doi.org/10.1103/PhysRevLett.74.1155} {\bibfield  {journal} {\bibinfo  {journal} {Phys. Rev. Lett.}\ }\textbf {\bibinfo {volume} {74}},\ \bibinfo {pages} {1155} (\bibinfo {year} {1995})}\BibitemShut {NoStop}%
\bibitem [{\citenamefont {Rohlfing}\ \emph {et~al.}(1996)\citenamefont {Rohlfing}, \citenamefont {Kr\"uger},\ and\ \citenamefont {Pollmann}}]{PhysRevB.54.13759}%
  \BibitemOpen
  \bibfield  {author} {\bibinfo {author} {\bibfnamefont {M.}~\bibnamefont {Rohlfing}}, \bibinfo {author} {\bibfnamefont {P.}~\bibnamefont {Kr\"uger}},\ and\ \bibinfo {author} {\bibfnamefont {J.}~\bibnamefont {Pollmann}},\ }\href {https://doi.org/10.1103/PhysRevB.54.13759} {\bibfield  {journal} {\bibinfo  {journal} {Phys. Rev. B}\ }\textbf {\bibinfo {volume} {54}},\ \bibinfo {pages} {13759} (\bibinfo {year} {1996})}\BibitemShut {NoStop}%
\bibitem [{\citenamefont {Nurhidayah}\ \emph {et~al.}(2026)\citenamefont {Nurhidayah}, \citenamefont {Fajariah}, \citenamefont {Purnawati}, \citenamefont {Santoso},\ and\ \citenamefont {Sholihun}}]{NURHIDAYAH2026101363}%
  \BibitemOpen
  \bibfield  {author} {\bibinfo {author} {\bibfnamefont {N.}~\bibnamefont {Nurhidayah}}, \bibinfo {author} {\bibfnamefont {N.}~\bibnamefont {Fajariah}}, \bibinfo {author} {\bibfnamefont {D.}~\bibnamefont {Purnawati}}, \bibinfo {author} {\bibfnamefont {I.}~\bibnamefont {Santoso}},\ and\ \bibinfo {author} {\bibfnamefont {S.}~\bibnamefont {Sholihun}},\ }\href {https://doi.org/https://doi.org/10.1016/j.nxmate.2025.101363} {\bibfield  {journal} {\bibinfo  {journal} {Next Materials}\ }\textbf {\bibinfo {volume} {10}},\ \bibinfo {pages} {101363} (\bibinfo {year} {2026})}\BibitemShut {NoStop}%
\bibitem [{\citenamefont {Schenk}\ \emph {et~al.}(2015)\citenamefont {Schenk}, \citenamefont {Tadich}, \citenamefont {Sear}, \citenamefont {O'Donnell}, \citenamefont {Ley}, \citenamefont {Stacey},\ and\ \citenamefont {Pakes}}]{10.1063/1.4921181}%
  \BibitemOpen
  \bibfield  {author} {\bibinfo {author} {\bibfnamefont {A.}~\bibnamefont {Schenk}}, \bibinfo {author} {\bibfnamefont {A.}~\bibnamefont {Tadich}}, \bibinfo {author} {\bibfnamefont {M.}~\bibnamefont {Sear}}, \bibinfo {author} {\bibfnamefont {K.~M.}\ \bibnamefont {O'Donnell}}, \bibinfo {author} {\bibfnamefont {L.}~\bibnamefont {Ley}}, \bibinfo {author} {\bibfnamefont {A.}~\bibnamefont {Stacey}},\ and\ \bibinfo {author} {\bibfnamefont {C.}~\bibnamefont {Pakes}},\ }\href {https://doi.org/10.1063/1.4921181} {\bibfield  {journal} {\bibinfo  {journal} {Applied Physics Letters}\ }\textbf {\bibinfo {volume} {106}},\ \bibinfo {pages} {191603} (\bibinfo {year} {2015})}\BibitemShut {NoStop}%
\bibitem [{\citenamefont {Schenk}\ \emph {et~al.}(2016)\citenamefont {Schenk}, \citenamefont {Tadich}, \citenamefont {Sear}, \citenamefont {Qi}, \citenamefont {Wee}, \citenamefont {Stacey},\ and\ \citenamefont {Pakes}}]{Schenk_2016}%
  \BibitemOpen
  \bibfield  {author} {\bibinfo {author} {\bibfnamefont {A.~K.}\ \bibnamefont {Schenk}}, \bibinfo {author} {\bibfnamefont {A.}~\bibnamefont {Tadich}}, \bibinfo {author} {\bibfnamefont {M.~J.}\ \bibnamefont {Sear}}, \bibinfo {author} {\bibfnamefont {D.}~\bibnamefont {Qi}}, \bibinfo {author} {\bibfnamefont {A.~T.~S.}\ \bibnamefont {Wee}}, \bibinfo {author} {\bibfnamefont {A.}~\bibnamefont {Stacey}},\ and\ \bibinfo {author} {\bibfnamefont {C.~I.}\ \bibnamefont {Pakes}},\ }\href {https://doi.org/10.1088/0957-4484/27/27/275201} {\bibfield  {journal} {\bibinfo  {journal} {Nanotechnology}\ }\textbf {\bibinfo {volume} {27}},\ \bibinfo {pages} {275201} (\bibinfo {year} {2016})}\BibitemShut {NoStop}%
\bibitem [{\citenamefont {Gomez}\ \emph {et~al.}(2024)\citenamefont {Gomez}, \citenamefont {Cruz}, \citenamefont {Milne}, \citenamefont {Debnath}, \citenamefont {Birdwell}, \citenamefont {Garratt}, \citenamefont {Pate}, \citenamefont {Rudin}, \citenamefont {Ruzmetov}, \citenamefont {Weil}, \citenamefont {Shah}, \citenamefont {Ivanov}, \citenamefont {Lake}, \citenamefont {Groves},\ and\ \citenamefont {Neupane}}]{10.1063/5.0203185}%
  \BibitemOpen
  \bibfield  {author} {\bibinfo {author} {\bibfnamefont {J.}~\bibnamefont {Gomez}, \bibfnamefont {H.}}, \bibinfo {author} {\bibfnamefont {J.}~\bibnamefont {Cruz}}, \bibinfo {author} {\bibfnamefont {C.}~\bibnamefont {Milne}}, \bibinfo {author} {\bibfnamefont {T.}~\bibnamefont {Debnath}}, \bibinfo {author} {\bibfnamefont {A.~G.}\ \bibnamefont {Birdwell}}, \bibinfo {author} {\bibfnamefont {E.~J.}\ \bibnamefont {Garratt}}, \bibinfo {author} {\bibfnamefont {B.~B.}\ \bibnamefont {Pate}}, \bibinfo {author} {\bibfnamefont {S.}~\bibnamefont {Rudin}}, \bibinfo {author} {\bibfnamefont {D.~A.}\ \bibnamefont {Ruzmetov}}, \bibinfo {author} {\bibfnamefont {J.~D.}\ \bibnamefont {Weil}}, \bibinfo {author} {\bibfnamefont {P.~B.}\ \bibnamefont {Shah}}, \bibinfo {author} {\bibfnamefont {T.~G.}\ \bibnamefont {Ivanov}}, \bibinfo {author} {\bibfnamefont {R.~K.}\ \bibnamefont {Lake}}, \bibinfo {author} {\bibfnamefont {M.~N.}\ \bibnamefont {Groves}},\ and\ \bibinfo {author} {\bibfnamefont {M.~R.}\ \bibnamefont {Neupane}},\ }\href
  {https://doi.org/10.1063/5.0203185} {\bibfield  {journal} {\bibinfo  {journal} {The Journal of Chemical Physics}\ }\textbf {\bibinfo {volume} {161}},\ \bibinfo {pages} {064702} (\bibinfo {year} {2024})}\BibitemShut {NoStop}%
\end{thebibliography}%
    \vspace{2cm}
    \begin{flushright}
    \end{flushright}

\clearpage

%%%%%%%%%%%%%%%%%%%%%%%%%%%%%%%%%%%%%%%%%
% Supplemental Information
%%%%%%%%%%%%%%%%%%%%%%%%%%%%%%%%%%%%%%%%%

\setcounter{equation}{0}
\setcounter{figure}{0}
\setcounter{table}{0}
\setcounter{page}{1}
\setcounter{section}{0}
\makeatletter
\renewcommand{\theequation}{S\arabic{equation}}
\renewcommand{\thefigure}{S\arabic{figure}}
\renewcommand{\thetable}{S\arabic{table}}
\renewcommand{\thesection}{S\arabic{section}}
\renewcommand{\thepage}{S\arabic{page}}
\renewcommand{\appendixname}{}

\def\PATHFIG_supp{./}

\onecolumngrid

%%%%%%%%%%%%%%%%%%%%%%%%%%%%
\section*{Supplemental Materials}
%%%%%%%%%%%%%%%%%%%%%%%%%%%%

This Supplemental Material provides additional analyses supporting the structural and electronic properties discussed in the main text.
Figures S1 and S2 show the closest Wannier functions associated with the Si atom and the subsurface C atom in the Si square lattice on the diamond (001) surface, respectively.
Figure S3 presents the orbital-projected spectral weights of the Si square lattice,
resolved into the contributions from the Si atom and the C atom directly beneath it.
Figure S4 shows the optimized structure and atom-projected band structure of the clean Si(001) surface for comparison with the Si-based structures formed on the diamond (001) surface.

    \begin{figure*}[htbp]
        \centering
            \centering
            \includegraphics[width=\textwidth]{\PATHFIG_supp/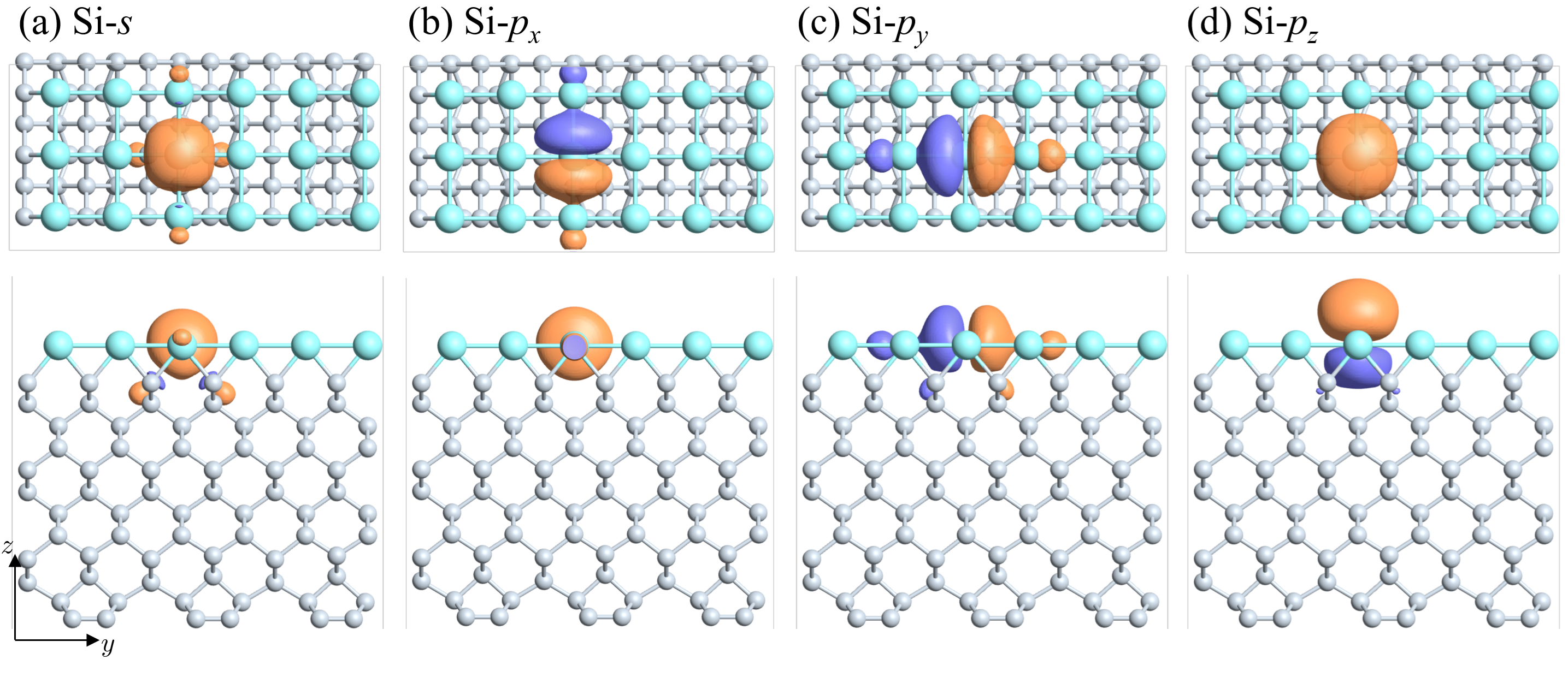}
            \caption{CWF of Si atom in Si square lattice on diamond (001) surface. The isosurface is plotted with isovalues of $\pm$0.03.}
            \label{fig:CWF_Si}
    \end{figure*}
    \begin{figure*}[htbp]
            \centering
            \includegraphics[width=\textwidth]{\PATHFIG_supp/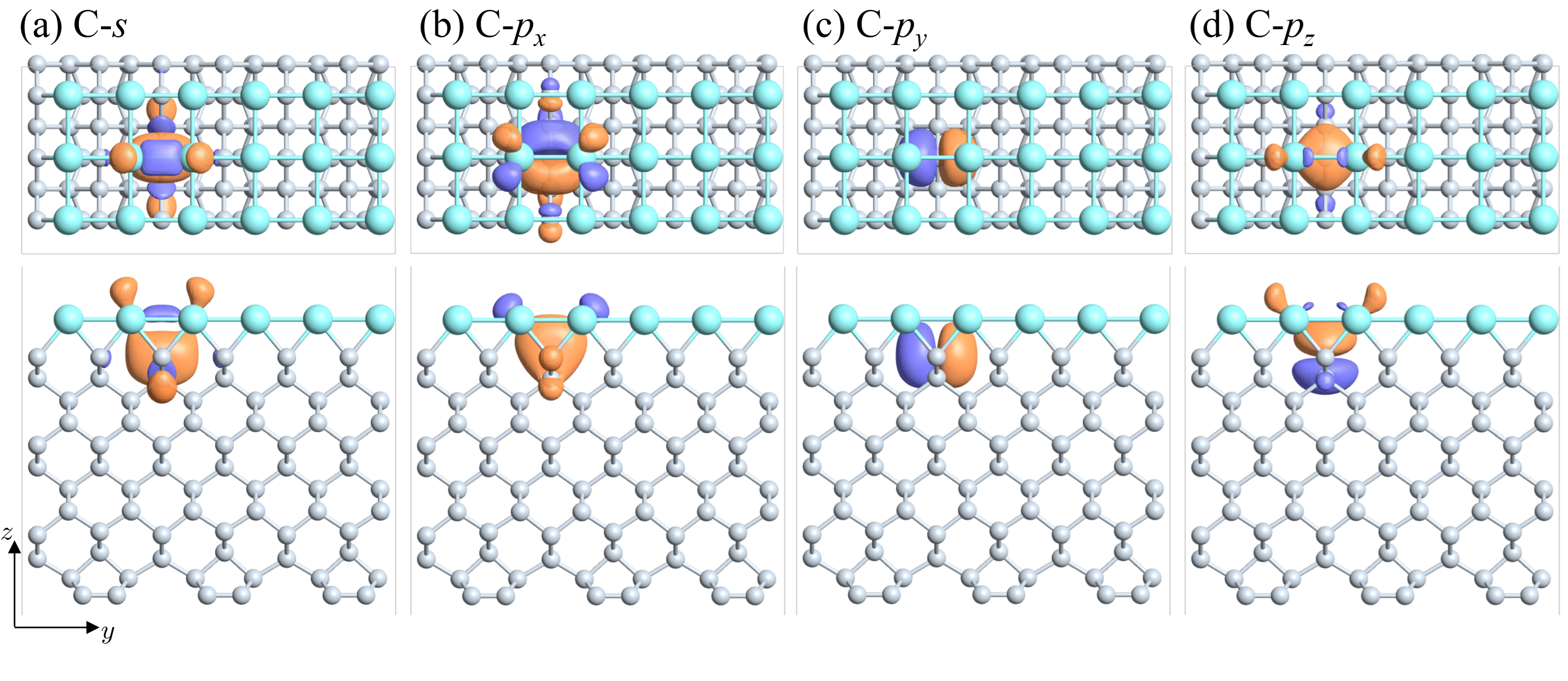}
            \caption{CWF of the subsurface C atom directly beneath the Si atom in Si square lattice on diamond (001) surface. The isosurface is plotted with isovalues of $\pm$0.03.}
            \label{fig:CWF_C}
    \end{figure*}
    
    \begin{figure*}[htbp]
            \centering
            \includegraphics[width=\textwidth]{\PATHFIG_supp/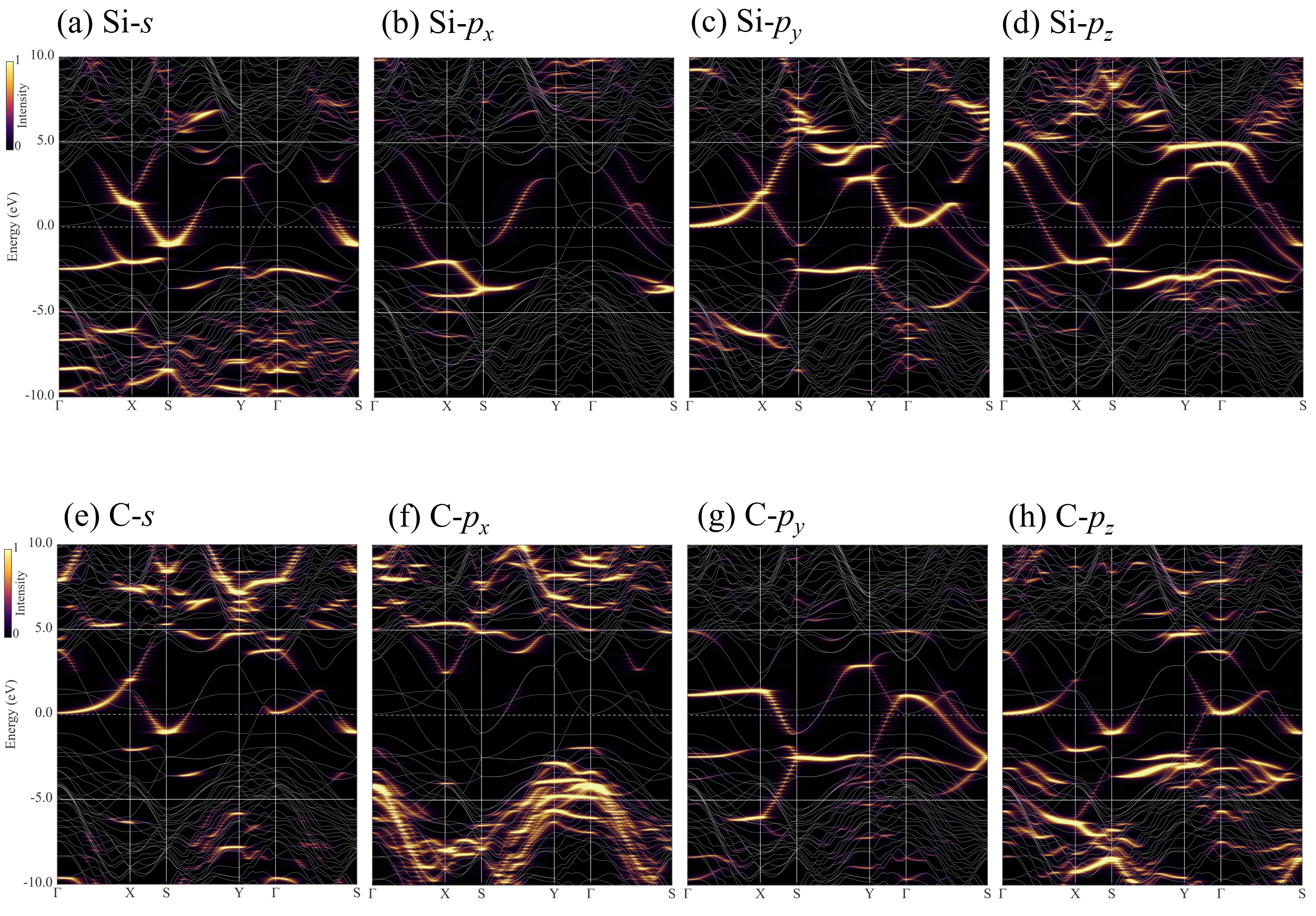}
            \caption{The orbital-projected spectral weights of the Si square lattice are shown for (a-d) the orbitals of a Si atom and (e-h) the orbitals of C atom below the adsorbed Si atom.}
            \label{fig:band_orb_si_square_lattice}
    \end{figure*}
    
    \begin{figure*}[htbp]
            \centering
            \includegraphics[width=\textwidth]{\PATHFIG_supp/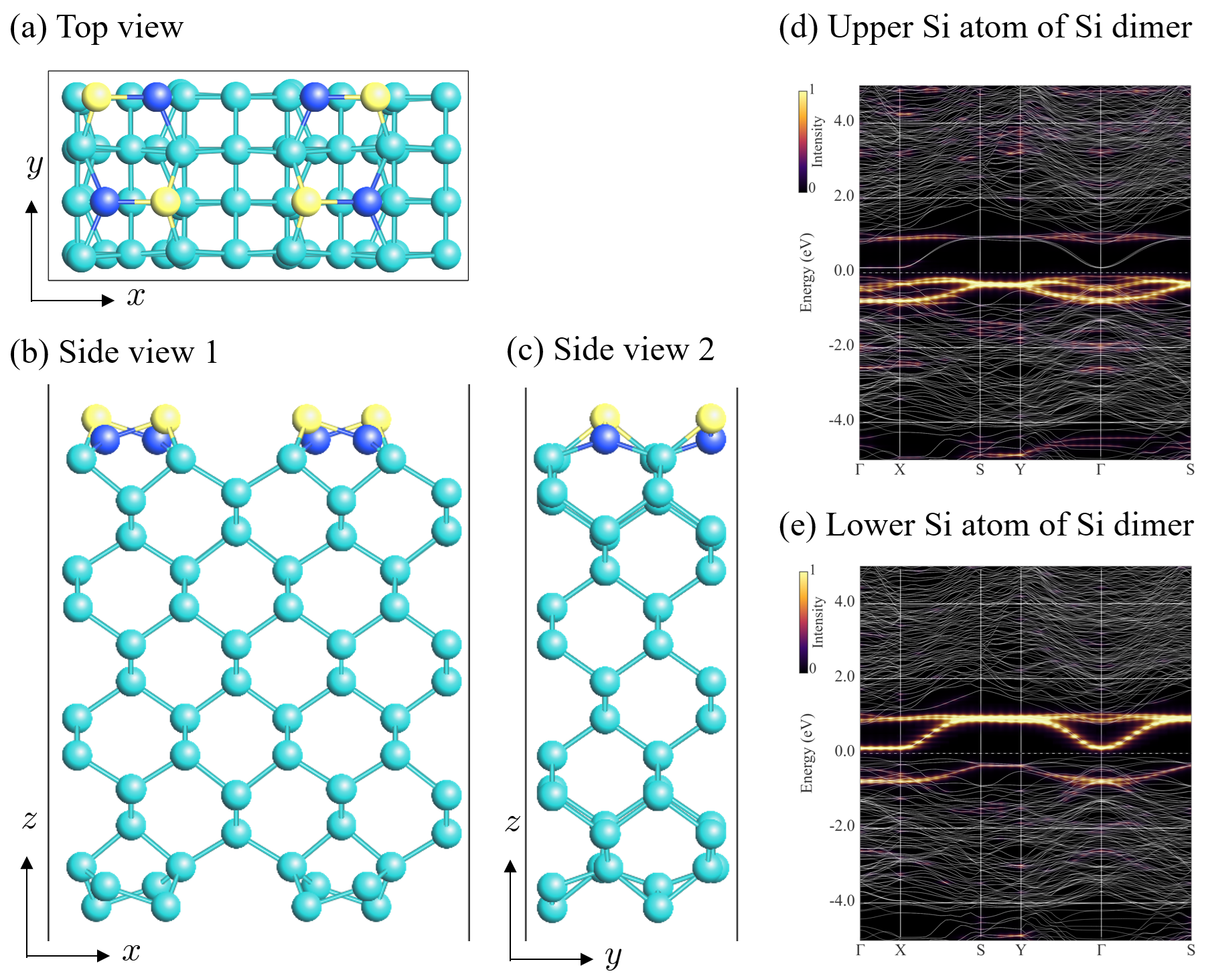}
            \caption{Optimized structure and band structure analysis of the Si(001) surface.
            (a) Top view and (b, c) side views of the optimized Si(001) surface.
            The atom-projected spectral weights are shown for (d) the upper Si atom of the Si dimer (yellow sphere) and (e) the lower Si atom of the Si dimer (blue sphere).}
            \label{fig:band_si_001}
    \end{figure*}
    
    %\begin{figure*}[htbp]
    %        \centering
    %        \includegraphics[width=\textwidth]{\PATHFIG_supp/stripe_df_all.pdf}
    %        \caption{Dielectric function of stripe structure on diamond (001) surface.}
    %        \label{fig:df_stripe}
    %\end{figure*}
    %\begin{figure*}[htbp]
    %        \centering
    %        \includegraphics[width=\textwidth]{\PATHFIG_supp/stripe_oc_all.pdf}
    %        \caption{Optical conductivity of stripe structure on diamond (001) surface.}
    %        \label{fig:oc_stripe}
    %\end{figure*}
    %\begin{figure*}[htbp]
    %\centering
    %\includegraphics[width=\textwidth]{\PATHFIG_supp/square_df_all.pdf}
    %\caption{Dielectric function of Si square lattice on diamond (001) surface.}
    %\label{fig:df_square}
    %\end{figure*}
    %\begin{figure*}[htbp]
    %        \centering
    %        \includegraphics[width=\textwidth]{\PATHFIG_supp/square_oc_all.pdf}
    %        \caption{Optical conductivity of Si square lattice on diamond (001) surface.}
    %        \label{fig:oc_square}
    %\end{figure*}
    %\begin{figure}[h]
    %    \centering
    %    \includegraphics[width=0.5\textwidth]{Images/Metastable xz.png}
    %    \caption{Metastable surface configuration of Si on C(001) surface}
    %    \label{fig:metastable_configuration}
    %\end{figure}
\end{document}